\documentclass[12pt]{iopart}
\usepackage{iopams}
\pdfoutput=1
\expandafter\let\csname equation*\endcsname=\relax
\expandafter\let\csname endequation*\endcsname=\relax
\usepackage{color}
\usepackage[breakwords]{truncate}
\usepackage[font={small}]{caption}
\usepackage{physics}
\usepackage{graphicx}
\usepackage{cite} 
\usepackage{subcaption}
\usepackage{etoolbox}
\usepackage[section]{placeins}
\usepackage[sectionbib]{chapterbib}
\usepackage{multirow}
\usepackage{pbox}
\usepackage{calc}
\usepackage{braket}
\usepackage[extra]{tipa}
\usepackage{csquotes}
\usepackage[breaklinks,colorlinks = true,urlcolor=blue,citecolor=blue]{hyperref}
\makeatletter
\renewcommand\@appendixstar{\@@par
 \ifnumbysec 
 \@addtoreset{table}{section}
 \@addtoreset{figure}{section}\fi
 \setcounter{section}{0}
 \setcounter{subsection}{0}
 \setcounter{subsubsection}{0}
 \setcounter{equation}{0}
 \setcounter{figure}{0}
 \setcounter{table}{0}
  \def\thesection{\Alph{section}} 
 \def\theequation{\ifnumbysec
      \Alph{section}.\arabic{equation}\else
      \Alph{section}\arabic{equation}\fi}
 \def\thetable{\ifnumbysec
      \Alph{section}\arabic{table}\else
      A\arabic{table}\fi}
 \def\thefigure{\ifnumbysec
      \Alph{section}\arabic{figure}\else
      A\arabic{figure}\fi}}
\makeatother

\begin{document}
\newcommand{\ben}[1]{\textcolor{blue}{\textbf{#1}}}
\title[First passage statistics of active random walks]{First passage statistics of active random walks on one  and two dimensional lattices} 

\author{Stephy Jose}
\address{Tata Institute of Fundamental Research, Hyderabad, India, 500046}
\eads{\mailto{stephyjose@tifrh.res.in}}
\date{\today}

\begin{abstract}
We investigate the first passage statistics of active continuous time random walks with Poissonian waiting time distribution on a one dimensional infinite lattice and a two dimensional infinite square lattice. We study the small and large time properties of the probability of the first return to the origin as well as the probability of the first passage to an arbitrary lattice site. It is well known that the occupation probabilities of an active particle resemble that of an ordinary Brownian motion with an effective diffusion constant at large times. Interestingly, we demonstrate that even at the leading order, the first passage probabilities are not given by a simple effective diffusion constant. We demonstrate that at late times, activity enhances the probability of the first return to the origin and the probabilities of the first passage to lattice sites close enough to the origin, which we quantify in terms of the P\'eclet number. Additionally, we derive the first passage probabilities of a symmetric random walker and a biased random walker without activity as limiting cases. We verify our analytic results by performing kinetic Monte Carlo simulations of an active random walker in one and two dimensions.
\end{abstract}


\newpage
{\pagestyle{plain}
 \tableofcontents
\cleardoublepage}

\section{Introduction}
\label{intro}

Active matter~\cite{czirok2000collective,toner2005hydrodynamics,bricard2013emergence,cates2012diffusive,cates2015motility} consists of constituent particles that move by self-propulsion and perform directed motion.  These systems violate detailed balance at the microscopic scale and serve as paradigmatic non-equilibrium systems to model a wide variety of phenomena in nature such as flocking, collective motion, non-equilibrium phase separation, and pattern formation~\cite{wysocki2014cooperative,stenhammar2014phase,speck2014effective,yang2014aggregation,fily2014freezing,enculescu2011active,lee2013active}.~Active particle models were initially introduced to study self-propelling particles such as bacteria~\cite{berg2004coli,powers2002role,locsei2007persistence,tailleur2008statistical,paoluzzi2013effective,di2010bacterial,saragosti2011directional} where the overall motion of a bacterium is described in terms of alternating run phases of directed motion of near-constant speed along a favorable direction set by an internal bias and tumbles where the particle changes its direction. Two major active particle models that have recently been of broad interest are the run and tumble particle (RTP) model~\cite{malakar2018steady,evans2018run,mori2020universal,mori2020universalp,singh2019generalised,angelani2014first,martens2012probability,slowman2017exact} and the active Brownian particle (ABP) model~\cite{stenhammar2013continuum,basu2018active,lindner2008diffusion,kumar2020active,romanczuk2012active,romanczuk2010collective}. The microscopic dynamics of a particle in both these models can be described by the same Langevin equations, but the direction flips stochastically at a fixed rate ($\gamma$) in the RTP model whereas the direction changes gradually by rotational diffusivity ($D_r$) in the ABP model. Apart from bacterial motion, these models also represent a wide range of disparate systems like animal groups, human crowds, liquid crystals, and colloids, amongst others~\cite{decamp2015orientational,redner2013reentrant,mallory2014anomalous,walsh2017noise,schnitzer1993theory,gautrais2009analyzing,cavagna2010scale,ramaswamy2010mechanics}.

The fundamental question of first passage to an arbitrary point has long attracted people's interest due to its applications in various disciplines such as finance, biology, search processes and chemical reactions~\cite{chou2014first,kenwright2012first,condamin2008probing,szabo1980first,park2003reaction,liu2017anchoring,zhang2009first,chicheportiche2014some}. Starting from discrete time random walks, first passage studies in stochastic processes have a and long rich history~\cite{siegert1951first,polya1921aufgabe,montroll1965random,montroll1969random,domb1954multiple,lindenberg1980lattice,burkhardt2014first,benichou2014first,khantha1983first,balakrishnan1983first,redner2001guide,majumdar1999persistence,bray2013persistence}. There is also abundant literature on continuous time random walks (CTRWs)~\cite{montroll1973random,montroll1979enriched,shlesinger1974asymptotic,klafter1980derivation,metzler2000random,bouchaud1990anomalous,bel2005occupation,kutner2017continuous,mainardi2020advent}, where the jumps are drawn from arbitrary waiting time distributions. In this paper, we compute the first passage statistics of an active continuous time random walker with nearest neighbor jumps on a one dimensional infinite lattice and a two dimensional infinite square lattice. For this, we study the model of active random walks introduced in \cite{PhysRevE.105.064103}. In this model, we consider the motion of a single active random walker starting from the origin at time $t=0$, which randomly changes its orientation along the lattice directions. Although a large literature is devoted to arbitrary waiting time distributions which in turn can lead to anomalous diffusion~\cite{bouchaud1990anomalous,metzler2000random,bel2005occupation}, we consider the simple case where the time gap between events are Poisson distributed. We study the probability of the first return to the origin, as well as the probability of the first passage to any arbitrary site other than the origin for an active random walker in one and two dimensions. For this, we use the fundamental recursion relations~\cite{siegert1951first,haus1987diffusion,redner2001guide,foong1994properties} connecting the characteristic functions of the occupation probabilities and the first passage probability distributions.

We summarize the main results of this paper below:

(a)~It is well known that the motion of self propelled particles can be described by an effective diffusion constant ${\mathcal{D}}$, at large times~\cite{howse2007self,lindner2008diffusion,solon2015active,aragones2018diffusion,malakar2018steady}. Consistent with the previous findings in the literature, we demonstrate that at large times, the occupation probability of an active random walker resembles that of a symmetric random walker with a modified diffusion constant at the leading order. However, we show that this result is modified when one considers the \textit{corrections} at sub-leading orders. In particular, using the exact expression of the characteristic function of the occupation probabilities, we show that active particle motion is not governed by the same modified diffusion constant~$\mathcal{D}$, at intermediate times. 

(b)~The first passage probabilities of active particles in continuous space have been explored previously mostly by using Fokker-Planck equations for survival probabilities~\cite{malakar2018steady,basu2018active,angelani2014first} or by using discrete time models~\cite{lacroix2020universal}. The survival probability $S(x,t)$ can be defined as the probability that a RTP starting from a location $x \ge 0$ does not cross the origin until time $t$. However, since the boundary condition $S(0,t)$, at the origin is unspecified, this approach cannot be directly used to investigate the probability of the first return to the origin. In this paper, we compute the probability of the first return to the origin by applying the recursion relation connecting the occupation probability of the origin and the first return probability to the origin~\cite{haus1987diffusion}. Surprisingly, we show that at large times, even to the leading order, the first return probability of an active random walker cannot be readily derived from a Brownian motion with a modified diffusion constant ${\mathcal{D}}$, in both one and two dimensions. We demonstrate that at large times, activity enhances the first return probabilities, which we quantify in terms of the P\'eclet number~\cite{ray2019peclet,kourbane2018exact}.

(c)~We also study the probability of the first passage to an arbitrary site different from the origin. The first passage properties of a diffusive RTP in one dimension in continuous space have been previously analyzed in \cite{malakar2018steady}, where the asymptotic behavior of the probability of the first passage to a location $x$, in the scaling limit~$x \rightarrow \infty,~t \rightarrow \infty$ keeping $x/t^{\frac{1}{2}}$ fixed was derived. In this scaling limit, the ﬁrst-passage time density reduces to that of an ordinary
Brownian motion with a modified diffusion constant $\mathcal {D}$. In this paper, we derive the exact asymptotic behavior of the probability of the first passage to any arbitrary lattice site ($t \rightarrow \infty$ limit with $x$ held fixed). For an active random walk in one dimension, we show analytically that at large times, the probability of the first passage to a lattice site  $x$, far enough from the origin can be considered to arise from a Brownian motion with an effective diffusion constant $\mathcal D$, but with a constant correction at the leading order. This correction, however, is prominently noticeable for any lattice site $x$, close enough to the origin. We also demonstrate that at large times, activity increases the likelihood of the first passage to lattice sites close enough to the origin. On the other hand, for lattice sites far enough from the origin, activity reduces the first passage probabilities at large times. Analogous to the one dimensional case, we also perform the same calculations for an active random walker on a two dimensional square lattice. For the two dimensional case, we numerically verify the departure of the first passage probabilities from the effective Brownian behavior at large times.

(d)~We find excellent agreement between our analytic results and kinetic Monte Carlo simulations of an active random walker on one  and two dimensional lattices. We also perform Monte Carlo simulations in continuous space and demonstrate that the qualitative predictions for the asymptotic limit of the first passage probabilities are the same in both cases.

(e)~Additionally, we check that the expressions for the occupation probabilities and the first passage probabilities of an active random walker reduce to that of a symmetric random walker and a biased random walker in the limiting cases. In one dimension, we derive the exact distribution for the first return probability of a passive random walker in terms of a generalized hypergeometric function. 

This paper is organized as follows.
In section~\ref{sec:model}, we discuss the model and the fundamental equations used in the study. We investigate the small and large time behavior of the occupation probability of active random walks in one dimension in section~\ref{subsec:occ_prob_1d}. We study the different temporal aspects of the first passage distribution of active random walks in one dimension in section~\ref{subsec:first_passage_1d}. The results on the occupation probability and the first passage probability density for the two dimensional case are presented in sections~\ref{subsec:occ_prob_2d} and~\ref{subsec:first_passage_2d} respectively. We summarize the conclusions of the study in section~\ref{sec:conclusions}.
\section{Model and evolution equations}
\label{sec:model}
\subsection{Model}
We consider the motion of a single active random walker which can perform directed motion along any of the lattice orientations on a one dimensional infinite lattice and a two dimensional infinite square lattice. This model was recently introduced and studied in \cite{PhysRevE.105.064103}. The motion starts from the origin at time $t=0$. We study the simple case where the particle is allowed to jump only to any of the nearest neighbor sites. We associate an internal spin state $m$ to the particle representing the bias direction. Here, $m$  can take values $0$ or $1$ corresponding to the bias directions $x$ and $-x$ respectively in one dimension and $0,~1,~2$ or $3$ corresponding to the bias directions $x,~y,~-x$ and $-y$ respectively in two dimensions. We consider symmetric initial (time, $t=0$) conditions where the particle has equal initial probabilities to be in any of the possible internal states. The dynamics proceeds in continuous time and is described by the diffusion rate (denoted as $D_{1d}$ in one dimension and $D_{2d}$ in two dimensions), bias rate $\epsilon$ and flipping rate $\gamma$. The time interval between consecutive events (could be a jump to a nearest neighbor site or a change of the internal state) are Poisson distributed. 

\begin{figure} [t!]
\centering
 \includegraphics[width=0.97\linewidth]{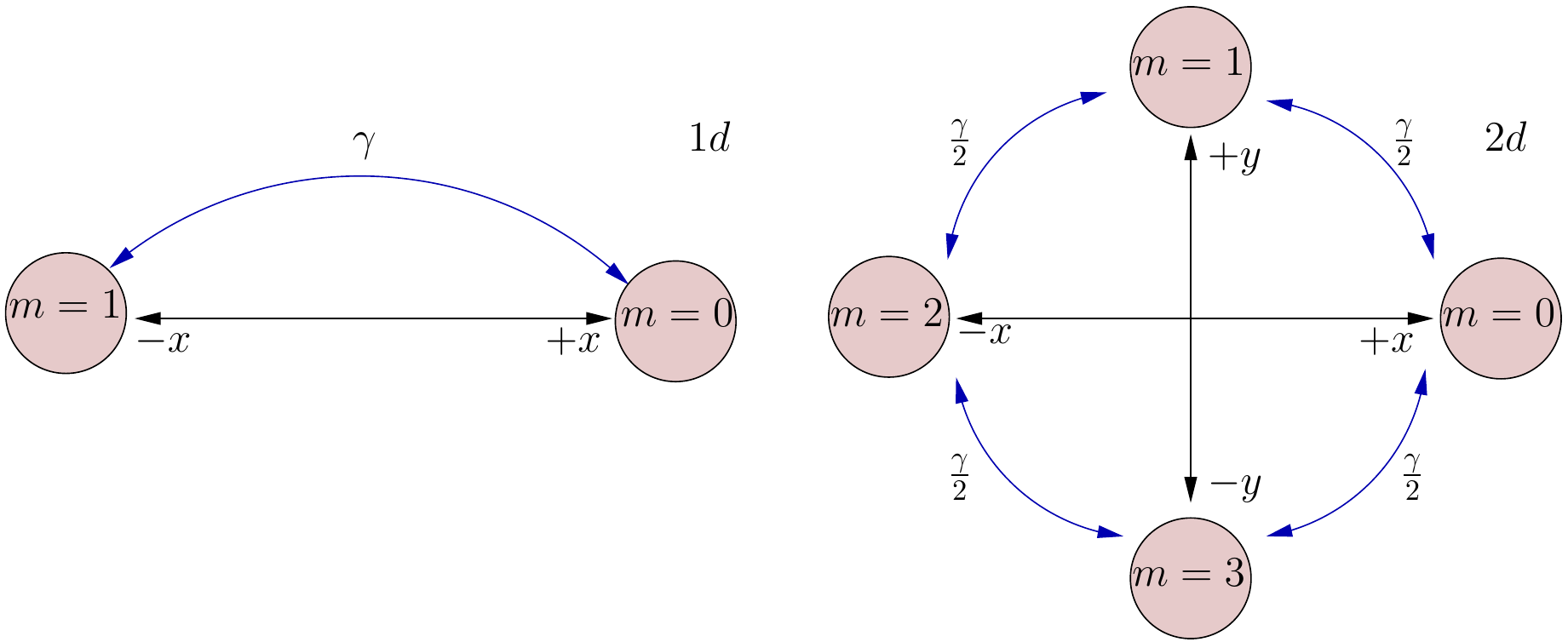}
 \caption{Active random walk model in one and two dimensions representing the states and the associated flipping rates. }\label{fig_model}
\end{figure}

In one dimension, an active particle in the internal state $m=0$ can make a hop in the $+x$ direction with a rate $D_{1d}+\epsilon$ and the $-x$ direction with a rate  $D_{1d}-\epsilon$. For the state $m=1$, the corresponding rates for both directions are reversed. The bias rate $\epsilon$, can take values between $0$ and $D_{1d}$. Along with translations, an active particle in the internal state $m$ can flip to the other state $m'$ with a rate $\gamma$. Thus the total rate for the active Poisson process in one dimension is $2D_{1d} + \gamma$. In two dimensions, an active particle in the internal state $m=0$ can make a hop in the $+x$ direction with a rate $D_{2d}+\epsilon$, $+y$ direction with a rate $D_{2d}$, $-x$ direction with a rate $D_{2d}-\epsilon$ and $-y$ direction with a rate $D_{2d}$. Similarly, rates are assigned to the states $1,~2$ and $3$ such that the particle is more biased along $+y,~-x$, and $-y$ directions respectively. The bias rate $\epsilon$, can take values between $0$ and $D_{2d}$. Along with translations, the particle in the internal state $m$ can flip its state to the other two possible internal states $m_{+1}$ and $m_{-1}$ with rates $\frac{\gamma}{2}$ each (refer to figure~\ref{fig_model}). Thus the total rate for the active Poisson process in two dimensions is $4D_{2d} + \gamma$.  
\subsection{Evolution equations}
In terms of discrete derivative operators, the evolution equations for the probability of occupation of site $\vec{r}$ at time $t$ denoted as $P_m(\vec{r},t) \equiv P_m$, of the particle in the internal state $m$ for the one  and two dimensional cases can be written as
\begin{equation}
\label{sq-1}
1d:~~\frac{\partial P_{m}}{\partial t} = D_{1d} {\nabla}^2 P_{m} -  \epsilon {\bf \hat{m}} . \vec{\nabla}P_{m} +\gamma \left( P_{m'}-P_{m}\right),
\end{equation}
\begin{equation}
\label{sq0}
2d:~~\frac{\partial P_{m}}{\partial t} = D_{2d} {\nabla}^2 P_{m} -  \epsilon {\bf \hat{m}} . \vec{\nabla}P_{m} +\frac{\gamma}{2}\left( P_{m_{+1}} + P_{m_{-1}}-2 P_{m}\right).
\end{equation}
In the above equations, $\vec{\nabla}$ is the discrete derivative operator and ${\nabla}^2$ is the discrete Laplacian operator. Also, $m'=\mod\left(m+1,2\right)$,~$m_{+1}=\mod\left(m+5,4\right)$ and $m_{-1}=\mod\left(m+3,4\right)$. For example in two dimensions, the particle in the internal state $m=0$ can flip to the state $m_{+1}=1$ or $m_{-1}=3$ with a rate $\frac{\gamma}{2}$.
Also, ${\bf \hat{m}}$ denotes the bias direction for the particle in the internal state $m$. The total probability for the particle to be at a lattice site $\vec{r}$ at time $t$ in either of the states is given as $P(\vec{r},t)=\sum_m P_m(\vec{r},t)$.

Another crucial observable of interest in random walks is the probability of the first passage to an arbitrary lattice site.
The occupation probabilities and the first passage probabilities  are directly linked through recursion relations in Laplace space. In this paper, we derive the exact asymptotic behavior of the probability of the first return to the origin, as well as the probability of the first passage to any arbitrary lattice site. We compute the first passage probability density $F(\vec{r},t)$, defined as the probability per unit time that a random walker starting from the origin at time $t=0$ arrives at the lattice location $\vec{r}$ for the first time at time $t$. By definition, $F(\vec{r},0)=0$ for all lattice sites including the origin. The first passage probability density $F(\vec{r},t)$ and the occupation probability $P(\vec{r},t)$ are related as~\cite{haus1987diffusion,redner2001guide,jose2021first}
\begin{equation}
\label{eq:a9ab}
P(\vec{r},t)=\int_{0}^{t}F(\vec{r},t')P(\vec{0},t-t')dt'+\delta_{\vec{r},\vec{0}}e^{-Wt},
\end{equation}
where $W=2D_{1d}$ for the one dimensional case and $W=4D_{2d}$ for the two dimensional case.
 The first term in the RHS of the above equation corresponds to walks that first arrive at the position $\vec{r}$ at time $t' \le t$ and then return to $\vec {r}$ in the remaining time $t-t'$ which is precisely the form of the convolution integral in equation~(\ref{eq:a9ab}). The second term corresponds to the initial condition and the waiting probability at the origin. The Laplace transform of $P(\vec{r},t)$ is defined as $\tilde P(\vec{r},s)=\int_{-\infty}^{\infty}dt e^{-st}P(\vec{r},t)$. Multiplying equation~(\ref{eq:a9ab}) throughout by $e^{-st}$ and integrating with respect to $t$ yields
\begin{equation}
\tilde F(\vec{r},s)=\frac{\tilde P(\vec{r},s)}{\tilde P(0,s)},~\forall~\vec{r} \neq \vec{0},
\label{reca0}
\end{equation}
and
\begin{equation}
\label{reca}
\tilde F(0,s)=1-\frac{1}{\tilde P(0,s)(s+W)},~\vec{r} = \vec{0},
\end{equation}
where $\tilde P(0,s)\equiv \tilde P(\vec{0},s)$ and $\tilde F(0,s)\equiv \tilde F(\vec{0},s)$.
These recurrence relations are extremely useful since, in most cases, the occupation probability is more readily available than the first passage time distribution, particularly in higher dimensions and for complex lattice structures. 
Finally, the recurrence properties of random walks can be investigated from the probability, $R(0,t)$ of returning to the origin up to time $t$, defined as
\begin{equation}
R(0,t)=\int_{0}^{t}{dt}^{'}F(0,{t}^{'}).
\end{equation}
The walk is said to be recurrent if the probability of ever returning to the origin is unity; i.e.,  $R(0,\infty)=1$.
\subsection{Limiting cases}
Throughout the calculations, we also investigate the limiting cases of a biased random walker (denoted with the subscript \enquote{$brw$}) by setting the flipping rate $\gamma=0$ and a symmetric random walker (denoted with the subscript \enquote{$srw$}) by setting the bias rate $\epsilon=0$ in the expressions for an active random walker. The limiting case of $\gamma=0$ is a special case where the expressions reduce to that of a biased random walker averaged over all biasing directions. The limiting case of $\epsilon=0$ yields the probability distributions for a symmetric random walker as the probability rates for all the states are the same and flipping states has no repercussions.
\section{Occupation probabilities of active random walks in one dimension}
\label{subsec:occ_prob_1d}

We first consider the motion of a single active random walker starting from the origin at time $t=0$ on a one dimensional infinite lattice. We assume symmetric initial conditions where the particle has equal initial probabilities (${1}/{2}$ each) to be in state $0$ or state $1$ at time $t=0$. The particle is weakly biased along the $+x$ direction or $-x$ direction if it is in state 0 or state 1 respectively. In equation~(\ref{sq-1}), we provide the evolution equation for the occupation probability $P_m \equiv P_m (x,t)$, of a lattice site $x$ by an active particle in the internal bias direction $m$. The total probability to occupy a lattice site $x$ at time $t$ in either of the states is given as $P(x,t)=\sum_{m=0}^1 P_m(x,t)$.

We begin by studying the Fourier-Laplace transform of the occupation probability of a lattice site $x$, in one dimension defined as $\tilde P (k,s)=\sum_{x=-\infty}^{\infty}\int_0^{\infty}dte^{ik x-s t}P(x,t)$. We derive a  closed-form expression for $\tilde P (k,s)$ by taking a Fourier-Laplace transform of equation~(\ref{sq-1}). The Fourier-Laplace transform of the total occupation probability $P(x,t)$ is given as $\tilde P(k,s)=\sum_{m=0}^1 \tilde P_m(k,s)$. After simplification, we obtain the expression for $\tilde P (k,s)$ as in \cite{PhysRevE.105.064103}
\begin{equation}
\label{fl}
\tilde P (k,s)=\frac{1}{(s+4D_{1d}\sin^2 \frac{k}{2})+\frac{4 \epsilon^2 \sin^2 k}{(s+4D_{1d}\sin^2 \frac{k}{2}+2 \gamma)}}.
\end{equation}
The Fourier transform in equation~(\ref{fl}) can be inverted as
\begin{equation}
\label{fl_inv}
\tilde P(x,s)=\frac{1}{2 \pi}\int_{-\pi}^{\pi}dk\frac{e^{-ikx}}{(s+4D_{1d}\sin^2 \frac{k}{2})+\frac{4 \epsilon^2 \sin^2 k}{(s+4D_{1d}\sin^2 \frac{k}{2}+2 \gamma)}}.
\end{equation}
The exact expression for the characteristic function of the occupation probability of a lattice site $x$, is obtained by evaluating the above integral. We consider the cases for $x=0$ and $x \ne 0$ separately as the fundamental equations connecting the characteristic functions of the first passage probability distributions and the occupation probabilities are different for both cases [refer to equations~(\ref{reca0}) and (\ref{reca})].
\subsection{For \texorpdfstring{$x=0$}{Lg}}
Using the substitution $z=e^{ik}$, the integral in equation~(\ref{fl_inv}) is converted to an integral over a unit circle in complex plane solving which we obtain the Laplace transform of the occupation probability of the origin $\tilde P(x=0,s)\equiv \tilde P(0,s)$ as
\begin{equation}
\label{rtp}
\tilde P(0,s)=
\sqrt{\frac{4 {D_{1d}}^2\gamma ^2+4 \epsilon ^2 \gamma  (s+2 D_{1d}) +2 \epsilon ^2 [\tilde h(s)+s (s+4 D_{1d})]}{s (s+4 D_{1d}) \tilde f(s)^2}}.
\end{equation}
The explicit forms of the functions  $\tilde f(s)$ and $\tilde h(s)$ are given as 
\begin{eqnarray} 
\tilde f(s)&=&
\sqrt{4 {D_{1d}}^2 \gamma ^2+4 \epsilon ^2 [2 \gamma (s+2 D_{1d})+s (s+4 D_{1d})]+16 \epsilon ^4},\nonumber\\
\tilde h(s)&=&\sqrt{s (s+4 D_{1d})(s+2 \gamma ) (s+4D_{1d}+2 \gamma)}.
\label{fshs}
\end{eqnarray} 
The details of the above calculation are provided in appendix~\ref{appendix_a0}.
The function $\tilde P(0,s)$ displays a divergence as $s \rightarrow 0$. Although it is hard to invert the Laplace transform in equation~(\ref{rtp}) exactly, it is possible to derive the limiting behaviors of the occupation probability of the origin from this exact expression.
\subsubsection{Limiting cases}

It is instructive to examine various limits of the exact expression for the Laplace transform of the occupation probability of the origin of an active random walker provided in equation~(\ref{rtp}). Setting $\gamma=0$ in equation~(\ref{rtp}) yields the biased random walk ($brw$) result
\begin{equation}
\label{0}
{\tilde P(0,s)}_{brw}=\frac{1}{\sqrt{s(s+4 D_{1d})+4 \epsilon^2}}.
\end{equation}
Similarly, substituting $\epsilon=0$ in equation~(\ref{rtp}) produces the symmetric random walk ($srw$) result
\begin{equation} 
\label{1}
{\tilde P(0,s)}_{srw}=\frac{1}{\sqrt{s (s+4 D_{1d})}}.
\end{equation}
These Laplace transforms can be easily inverted and we obtain the expression~\cite{feller2008introduction,balakrishnan1983some,jose2021first}
\begin{equation}
\label{eq:a3ab}
{P(0,t)}_{brw}=e^{-2 D_{1 d}t}I_0( 2\sqrt{\eta }~t),
\end{equation}
where
\begin{equation}
\eta ={D_{1d}}^2- \epsilon ^2,
\label{eta}
\end{equation}
and $I_n(z)$ is the modified Bessel function of order $n$.
For a symmetric random walk ($\epsilon=0$, $\eta ={D_{1d}}^2$), the expression in equation~(\ref{eq:a3ab}) reduces to 
\begin{equation}
\label{eq:a3b}
{P(0,t)}_{srw}=e^{-2 D_{1d}t}I_0(2 D_{1d}t).
\end{equation}

The main focus of this study of active random walk is to study the convergence to or deviation from a passive random walk in different limits.
We next analyze the small and large $s$ behaviors of equation~(\ref{rtp}) to extract the large and small time behaviors of the occupation probability of the origin, $P(0,t)$.
In the small and large $s$ limits, we expand the expression presented in equation~(\ref{rtp}) to obtain 
\begin{eqnarray} 
\label{p0sarw_series}
\hspace{-1 cm}
\tilde P(0,s)&\xrightarrow[s \rightarrow 0]{}&\tilde{\alpha}_{\frac{1}{2}} \sqrt{\pi} \frac{1}{s^{\frac{1}{2}}}+\tilde{\alpha}_{1} s^0-2 \tilde{\alpha}_{\frac{3}{2}} \sqrt{\pi} s^{\frac{1}{2}}+\tilde{\alpha}_{2}s^1+\frac{4}{3} \tilde{\alpha}_{\frac{5}{2}} \sqrt{\pi} s^{\frac{3}{2}}-...~,\nonumber\\\hspace{-1 cm}
\tilde P(0,s)&\xrightarrow[s \rightarrow \infty]{}& \tilde{\beta}_{0} \frac{1}{s}+\tilde{\beta}_{1} \frac{1}{s^2}+  2 \tilde{\beta}_{2} \frac{1}{s^3}+ 6 \tilde{\beta}_{3}\frac{1}{s^4}+...~.
\end{eqnarray}
We use the superscript tilde \enquote{$\sim$}, to denote the coefficients appearing in the expansions for an active random walker. The coefficients appearing in the corresponding expansions for a passive random walker are represented with the same symbols but without the superscript. The expansions provided in~equation~(\ref{p0sarw_series}) are valid only for non-zero flipping rate $\gamma$.
The coefficients appearing in the above expressions ($\{\tilde{ \alpha}_i \}$ and $\{ \tilde{\beta}_i \}$) can be determined precisely, and we present a list of the expressions for the first few coefficients in table~\ref{table_arw} of appendix~\ref{appendix_a}. The order in which the coefficients $ \tilde{\alpha}_i $ and $ \tilde{\beta_i }$ appear in the appropriate time domain expansions is indicated by the subscript $i$. As in equation~(\ref{p0sarw_series}), we obtain the following limiting behaviors for a symmetric random walk and a biased random walk
\begin{eqnarray} 
\label{p0ssrw_series}
\hspace{-1.0 cm}
{\tilde P(0,s)}_{srw}&\xrightarrow[s \rightarrow 0]{}&\alpha_{\frac{1}{2}} \sqrt{\pi} \frac{1}{s^{\frac{1}{2}}}-2 \alpha_{\frac{3}{2}} \sqrt{\pi} s^{\frac{1}{2}}+\frac{4}{3} \alpha_{\frac{5}{2}} \sqrt{\pi} s^{\frac{3}{2}}-...~,\nonumber\\\hspace{-1.0 cm}
{\tilde P(0,s)}_{brw}&\xrightarrow[s \rightarrow \infty]{}& \beta_{0} \frac{1}{s}+\beta_{1} \frac{1}{s^2}+  2 \beta_{2} \frac{1}{s^3}+ 6 \beta_{3}\frac{1}{s^4}+...~.
\end{eqnarray}
The above equations are obtained by performing series expansions of the expressions provided in~equations~(\ref{1}) and~(\ref{0}) respectively.
We provide a list of the first few non-zero coefficients ($\{ {\alpha_i}\}$ and $\{ {\beta_i}\}$) appearing in the above equations in table~\ref{table_brw} of appendix~\ref{appendix_a}. The coefficient ${\beta_i}$ is exactly equal to the coefficient $\tilde{\beta}_i$ for $i=0,~1,~2$.
Thus over short periods of time, 
the occupation probability of an active random walker exactly resembles a biased random walker. This corresponds to time scales $t<< \frac{1}{\gamma}$, where the walker has not flipped its direction. We next focus on the small $s$ behavior of the characteristic function of the occupation probability of the origin of an active random walker provided in equation~(\ref{p0sarw_series}). It is apparent that the leading order behavior can be interpreted as a diffusive process with a {\it modified} diffusion constant $\mathcal{D}_{1d}$ ~\cite{howse2007self,lindner2008diffusion,solon2015active,aragones2018diffusion,malakar2018steady,PhysRevE.105.064103}, and we identify $\alpha_{\frac{1}{2}}(\mathcal{D}_{1d})=\tilde{\alpha}_{\frac{1}{2}}$ with
 \begin{equation} 
\label{d1dn}
{\mathcal{D}}_{1d}=  D_{1d}+\frac{2\epsilon^2}{\gamma}.
 \end{equation}
Here, $\tilde{\alpha}_{\frac{1}{2}}$ and ${\alpha_{\frac{1}{2}}}$ are the coefficients appearing in the leading term of~equations~(\ref{p0sarw_series}) and~(\ref{p0ssrw_series}) respectively. 

These occupation probabilities in Laplace space are particularly useful because they can be used to determine the limiting behavior of the occupation probabilities in the time domain. The  occupation probability of the origin at large and small times can be extracted by performing a term by term Laplace inversion of the expressions provided in equation~(\ref{p0sarw_series}). For this, it is useful to recall the identity for inverse Laplace transforms, $L^{-1}\left [s^{-(a+1)} \right ]=t^a/\Gamma (a+1)$, where $\Gamma$ is the gamma function. This yields the small and large time behavior of the occupation probability of the origin as
\begin{eqnarray} 
\label{eqa10}
P(0,t)&\xrightarrow[t \rightarrow \infty]{}& \tilde{\alpha}_{\frac{1}{2}}\frac{1}{t^{\frac{1}{2}}}+\tilde{\alpha}_{\frac{3}{2}} \frac{1}{t^{\frac{3}{2}}}+\tilde{ \alpha}_{\frac{5}{2}}\frac{1}{t^{\frac{5}{2}}}+...~,\nonumber\\
P(0,t)&\xrightarrow[t \rightarrow 0]{}& \tilde{\beta}_{0}+\tilde{\beta}_{1} t+\tilde{\beta}_{2} t^2+\tilde{\beta}_{3} t^3+...~.
\end{eqnarray}
The terms appearing at $O(s^i)$ where $i$ is a non-negative integer, in the small $s$ expansion of $\tilde P(0,s)$ provided in equation~(\ref{p0sarw_series}) do not contribute to the asymptotic limit of $P(0,t)$ as the gamma function diverges at these points. 

Since the leading term in the asymptotic limit of the occupation probability of an active random walk can be understood as coming from a symmetric random walk with a modified diffusion constant, it is also instructive to analyze the subleading corrections to the occupation probability of the origin. For a symmetric random walk, all corrections to the occupation probability in the long time limit are functions of the same diffusion constant $ {D}_{1d}$ and can be represented as
  \begin{equation}
 \label{dif_prob}
 {P(0,t)}_{srw}\xrightarrow[t \rightarrow \infty]{} \alpha_\frac{1}{2}({{D}}_{1d})\frac{1}{t^{\frac{1}{2}}}+\alpha_\frac{3}{2}({{D}}_{1d})\frac{1}{t^{\frac{3}{2}}}+\alpha_\frac{5}{2}({{D}}_{1d})\frac{1}{t^{\frac{5}{2}}}+...~.
 \end{equation}
However, as the expressions for the collection of coefficients $\{\tilde {\alpha}_i\}$ show, a single modified diffusion constant $\mathcal D_{1d}$, does not capture the features of an active random walk at all time scales~(refer to table~\ref{table_arw} of appendix~\ref{appendix_a}). 
\begin{figure} [t!]
\centering
 \includegraphics[width=0.95\linewidth]{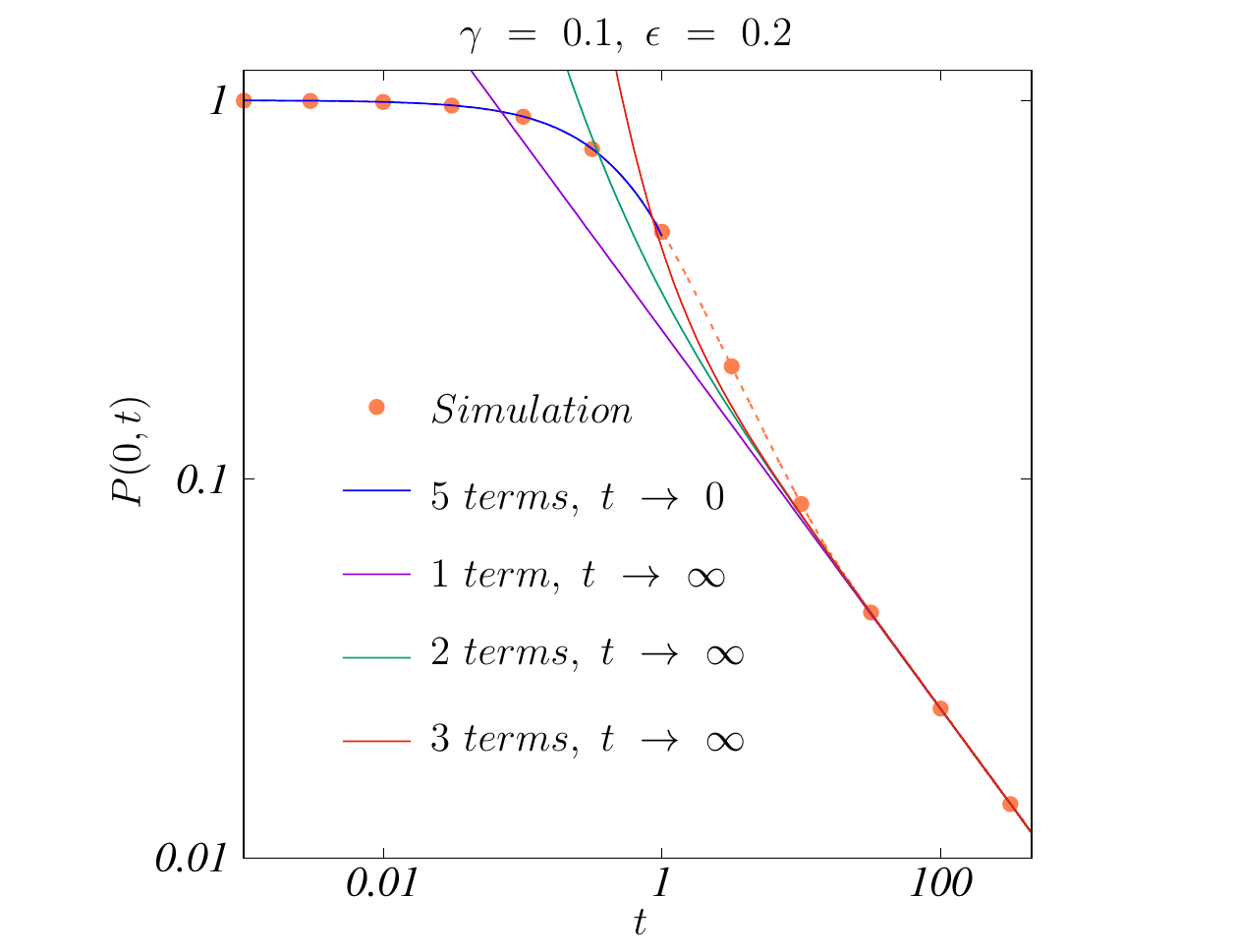}
 \caption{Continuous time kinetic Monte Carlo simulation results (points with the dashed line) for the occupation probability of the origin $P(0,t)$, of a RTP on a one dimensional infinite lattice plotted as a function of time for fixed parameter values $\gamma=0.1$, $\epsilon=0.2$ and $D_{1d}=0.5$. The solid curves correspond to the analytic expressions for the small and large time limits of $P(0,t)$ provided in equation~(\ref{eqa10}). As we keep more terms in the series expansions of $P(0,t)$, the analytic results converge to the simulation results.}\label{fig_p0t_comp}
\end{figure} 
We compare the analytic expressions for the large and small time behaviors  of the occupation probability of the origin of a RTP on a one dimensional lattice provided in equation~(\ref{eqa10}) with kinetic Monte Carlo simulation results in figure~\ref{fig_p0t_comp}. As we keep more terms in the series expansions provided in equation~(\ref{eqa10}), the analytic results converge to the simulation results. Having analyzed the occupation probability of the origin of an active random walker in one dimension, we next analyze the occupation probability of an arbitrary lattice site $x$, different from the origin.
\subsection{For \texorpdfstring{$x \ne 0$}{Lg}}
Similar to the case for $x=0$, we use the substitution $z=e^{ik}$ in equation~(\ref{fl_inv}) and solve the integral in complex space~(refer to appendix~\ref{appendix_a0}). Even though the integral can be computed exactly, the final expression obtained for the Laplace transform of the occupation probability of an arbitrary site $x$, denoted as $\tilde P (x,s)$ is rather long and we do not quote it here. Nevertheless, the long time behavior of the occupation probability $P(x,t)$ can be obtained by studying the small $s$ behavior of $\tilde P (x,s)$. For small $s$, we obtain
\begin{eqnarray} 
\label{series_pxs}
\hspace{-1.0 cm}
\tilde P(x,s)&\xrightarrow[s \rightarrow 0]{}&\exp\left(-\left| x\right| \sqrt{\frac{s}{\mathcal{D}_{1d}}} \right) \left [ \tilde{\sigma}_{\frac{1}{2}} \sqrt{\pi} \frac{1}{s^{\frac{1}{2}}}+\tilde{\sigma}_{1} s^0-2 \tilde{\sigma}_{\frac{3}{2}} \sqrt{\pi} s^{\frac{1}{2}}+...\right ].
\end{eqnarray}
Here, the set of coefficients $\{\tilde{\sigma}_{i}\}$ are functions of the rates $\gamma,~\epsilon$ and $D_{1d}$ and the variable $x$. We identify the coefficient $\tilde{\sigma}_{\frac{1}{2}}$ to be same as the coefficient $\tilde{\alpha}_{\frac{1}{2}}$ appearing in equation~(\ref{p0sarw_series}) while the rest are different. We provide explicit forms of the first few coefficients $\{\tilde{\sigma_{i}}\}$ in table~\ref{table_arw} of appendix~\ref{appendix_a}. We next derive the limiting behaviors of the occupation probability of an arbitrary lattice site for an active random walker in one dimension.
\subsubsection{Limiting cases}

We examine the various limiting cases of the occupation probability of an arbitrary lattice site $x \ne 0$, for an active random walker. Substituting $\gamma=0$ in the expression for the Laplace transform of the occupation probability of an active random walker yields the biased random walk ($brw$) result
\begin{equation}
\label{pxsbrw}
 {\tilde P(x,s)}_{brw}=
\frac{{\left(s+2 D_{1d}-\sqrt{s(s+4 D_{1d})+4 \epsilon^2}\right)}^{\left| x\right|}}{2^{\left| x\right|+1}\sqrt{s(s+4D_{1d})+4 \epsilon^2}}{\left[\frac{1}{{(D_{1d}- \epsilon)}^{\left| x\right|}}+\frac{1}{{(D_{1d}+ \epsilon)}^{\left| x\right|}}\right]}.
\end{equation} 

The above expression represents the characteristic function of the occupation probability of a biased random walker averaged over the two biasing directions. Similarly, substituting $\epsilon=0$ in the expression for the Laplace transform of the occupation probability of an active random walker yields the symmetric random walk $(srw)$ result
\begin{equation}
\label{pxssrw}
 {\tilde P(x,s)}_{srw}=\frac{\left(s+2 D_{1d}-\sqrt{s (s+4 D_{1d})}\right)^{\left| x\right| }}{(2 D_{1d})^{\left| x\right| } \sqrt{s (s+4 D_{1d})}}.
\end{equation} 
This is exactly the expression for the characteristic function of the occupation probability of a symmetric random walker. The expressions in~equations~(\ref{pxsbrw}) and~(\ref{pxssrw}) can be inverted exactly and we obtain the occupation probabilities in the time domain~\cite{feller2008introduction,balakrishnan1983some,jose2021first} as
\begin{equation}
\label{pxtbrw}
{P(x,t)}_{brw}=\frac{e^{-2D_{1d}t}}{2}I_{\left |x  \right |}\left(2 \sqrt{\eta}~t \right)\left [\sqrt{{\left(\frac{(D_{1d}+ \epsilon)}{(D_{1d}- \epsilon)}\right)}^{\left| x\right|}}+\sqrt{{\left(\frac{(D_{1d}- \epsilon)}{(D_{1d}+ \epsilon)}\right)}^{\left| x\right|}}~ \right ],
\end{equation}
where $\eta$ is defined in~equation~(\ref{eta}) and
\begin{equation}
\label{pxtsrw}
{P(x,t)}_{srw}=e^{-2 D_{1d} t} I_{\left| x\right|}\left(2 {{D_{1d}}} t\right).
\end{equation}
Since we are interested in studying the convergence of an active random walk to a symmetric random walk in the long time limit, we perform a series expansion of the expression provided in equation~(\ref{pxssrw}) around small $s$ to obtain
\begin{eqnarray} 
\label{series_pxs_srw}
\hspace{-1.5 cm}
{\tilde P(x,s)}_{srw}&\xrightarrow[s \rightarrow 0]{}&\exp\left(-\left| x\right| \sqrt{\frac{s}{{D}_{1d}}} \right) \left [ {\sigma_{\frac{1}{2}}} \sqrt{\pi} \frac{1}{s^{\frac{1}{2}}}-2 {\sigma_{\frac{3}{2}}} \sqrt{\pi} s^{\frac{1}{2}}+\sigma_2 s^1+...\right ].
\end{eqnarray}
We provide a list of the first few non-zero coefficients $\{ {\sigma_i}\}$, appearing in the above equation in table~\ref{table_brw} of appendix~\ref{appendix_a}. We identify $\sigma_{\frac{1}{2}}(\mathcal{D}_{1d})=\tilde{\sigma}_{\frac{1}{2}}$, where $\tilde{\sigma}_{\frac{1}{2}}$ and ${\sigma_{\frac{1}{2}}}$ are the coefficients appearing in the leading term of~equations~(\ref{series_pxs}) and~(\ref{series_pxs_srw}) respectively.
After Laplace inversion, we obtain the leading behavior of the occupation probability $P(x,t)$ of an active random walker in the asymptotic limit as
\begin{eqnarray} 
\label{series_pxt}
P(x,t)&\xrightarrow[t \rightarrow \infty]{}&\exp\left({\frac{-{\left| x\right|}^2}{4\mathcal{D}_{1d}t}} \right)  \frac{1}{\sqrt{4 \pi \mathcal{D}_{1d}t}}.
\end{eqnarray}
This expression resembles the occupation probability of an ordinary Brownian motion with a modified diffusion constant $ \mathcal{D}_{1d}$. However, the subleading corrections at intermediate times are not functions of the same diffusion constant $\mathcal{D}_{1d}$. Knowing the associated Laplace transforms, for example, can also be used to extract the limiting behaviors of other time dependent quantities such as the first passage probability distributions. Having analyzed the occupation probabilities, we next analyze the first passage probability distributions of an active random walker in one dimension.
\section{First passage statistics of active random walks in one dimension}
\label{subsec:first_passage_1d}
Since the characteristic functions of the occupation probabilities and the first passage probability densities are directly related through the renewal relations in equations~(\ref{reca0}) and (\ref{reca}), we apply the results for the occupation probabilities to analyze the first passage probability densities. As for the occupation probabilities, we address the scenarios for $x=0$ and $x \ne 0$ separately.
\subsection{For \texorpdfstring{$x = 0$}{Lg}}
We use the exact expression for the Laplace transform of the occupation probability of the origin $\tilde P(0,s)$, provided in equation~(\ref{rtp}) to obtain the expression for the Laplace transform of the first return probability density $\tilde F(0,s)$, of an active random walker on a one dimensional infinite lattice. 
Substituting equation~(\ref{rtp}) in  equation~(\ref{reca}) yields
\begin{equation}
\label{rtp2}
\tilde F(0,s)=
1-\sqrt{\frac{ s (s+4 D_{1d}) \tilde f(s)^2 /{(s+2 D_{1d})}^2 }{4 {D_{1d}}^2\gamma ^2+4 \gamma  (s+2 D_{1d}) \epsilon ^2+2 \epsilon ^2 [\tilde h(s)+s (s+4 D_{1d})]}},
\end{equation}
where the expressions for the functions $\tilde f(s)$, and ~$\tilde h(s)$ also appearing in equation~(\ref{rtp}) are provided in equation~(\ref{fshs}). We note that $\tilde F(0,0)=1$, implying that an active random walk is recurrent in one dimension. That is, an active random walker visits the origin infinitely often in an infinite time and the probability of ever returning to the origin is $1$. 
\subsubsection{Limiting cases}

We next analyze various limits of the characteristic function of the first return probability density provided in equation~(\ref{rtp2}).
Substituting $\gamma=0$ in equation~(\ref{rtp2}) yields the characteristic function of the first return probability density of a biased random walker ($brw$) as
\begin{equation}
\label{eq:a8}
{\tilde F (0,s)}_{brw}=1-\frac{\sqrt{s (s+4 D_{1d})+4 \epsilon ^2}}{(s+2 D_{1d})}.
\end{equation}
Similarly, we obtain the characteristic function of the first return probability density of a symmetric random walker ($srw$) by substituting the bias rate $\epsilon=0$ in equation~(\ref{rtp2}) as
\begin{equation}
\label{eq:a9}
{\tilde F (0,s)}_{srw}=1-\frac{\sqrt{s (s+4 D_{1d})}}{(s+2 D_{1d})}.
\end{equation}
The expression in equation~(\ref{eq:a9}) can also be obtained by setting $\epsilon=0$ in equation~(\ref{eq:a8}).
The probability of ever returning to the origin for a biased random walker can be found from the above equation by setting $s$ equal to zero and we obtain
\begin{equation}
R(0,\infty)_{brw}={\tilde F (0,0)}_{brw}=1- \frac{\epsilon}{D_{1d}} <1.
\end{equation}
From the above equation, it is clear that a biased random walk is transient in one dimension and the probability of ever returning to the origin is less than $1$. For a symmetric random walk with bias rate $\epsilon=0$, the walk is recurrent and the probability of ever returning to the origin is unity.
We invert the Laplace transform in equation~(\ref{eq:a8}) to obtain
\begin{equation}
\label{eq:a13}
{F(0,t)}_{brw}= 2{\eta } t e^{-2 D_{1d}t} \, _1F_2\left(\frac{1}{2};\frac{3}{2},2;\eta t^2 \right),
\end{equation}
where $\eta$ is defined in~equation~(\ref{eta}) and $\, _1F_2$ is the generalized hypergeometric function~\cite{bell2004special}.
Equation~(\ref{eq:a13}) is the exact expression for the first return time density of a biased random walker on a one dimensional infinite lattice. These kinds of generalized hypergeometric functions appear in various contexts of first passage problems related to random walks in different dimensions~\cite{hughes1995random}. An analogous expression for the first return probability has also been derived in~\cite{jose2021first} in terms of Struve and Bessel functions for a biased random walker in one dimension. 
For a symmetric random walk ( $\epsilon=0$, $\eta ={D_{1d}}^2$), the first return probability density reduces to
\begin{equation} 
\label{f0t_srw}
{F(0,t)}_{srw}= 2{{D_{1d}}^2} t e^{-2 D_{1d}t} \, _1F_2\left(\frac{1}{2};\frac{3}{2},2; {D_{1d}}^2{t}^2 \right).
\end{equation}

For an active random walk, we next study the convergence to and deviation from a passive random walk in terms of the first passage probabilities in different temporal limits.
Analogous to the calculations for the occupation probability, we perform a series expansion of the characteristic function of the first return probability density $\tilde F(0,s)$ provided in equation~(\ref{rtp2}) in the small and large $s$ limits to obtain
\begin{eqnarray}
\label{asym0n}
\tilde F(0,s)&\xrightarrow[s \rightarrow 0]{}& \tilde{\phi}_1s^0-2\tilde{\phi}_\frac{3}{2}\sqrt{\pi}s^{\frac{1}{2}}+\tilde{\phi}_2 s^1+\frac{4}{3}\tilde{\phi}_\frac{5}{2}\sqrt{\pi} s^{\frac{3}{2}}+...~,\nonumber\\
\tilde F(0,s)&\xrightarrow[s \rightarrow \infty]{}& \tilde{\rho}_1 \frac{1}{s^2}+2 \tilde{\rho}_2 \frac{1}{s^3}+6 \tilde{\rho}_3 \frac{1}{s^4}+...~.
\end{eqnarray}
The expressions for the first few coefficients ($\{ \tilde{\phi}_i \}$ and $\{ \tilde{\rho}_i \}$) appearing in the above equations are provided in table~\ref{table_arw} of appendix~\ref{appendix_a}. The index $i$ indicates the order in which the coefficients $ \tilde{\phi}_i $ and $ \tilde{\rho}_i $ appear in the corresponding expressions in the time domain. As in equation~(\ref{asym0n}), we obtain the following limiting behaviors for a symmetric random walk and a biased random walk
\begin{eqnarray} 
\label{f0ssrw_series}
\tilde F(0,s)_{srw}&\xrightarrow[s \rightarrow 0]{}& {\phi_1}s^0-2{\phi_\frac{3}{2}}\sqrt{\pi}s^{\frac{1}{2}}+\frac{4}{3}{\phi_\frac{5}{2}}\sqrt{\pi} s^{\frac{3}{2}}+...~,\nonumber\\
\tilde F(0,s)_{brw}&\xrightarrow[s \rightarrow \infty]{}& {\rho_1} \frac{1}{s^2}+2 {\rho_2} \frac{1}{s^3}+6 {\rho_3} \frac{1}{s^4}+...~.
\end{eqnarray}
The above equations are obtained by performing series expansions of the expressions provided in equations~(\ref{eq:a9}) and~(\ref{eq:a8}) respectively.
We provide a list of the first few non-zero coefficients ($\{ {\phi_i}\}$ and $\{ {\rho_i}\}$) appearing in the above equations in table~\ref{table_brw} of appendix~\ref{appendix_a}. The coefficients ${\phi_i}$ and $\rho_i$ are exactly equal to the coefficients $\tilde{\phi}_i$ and $\tilde{\rho}_i$ respectively for $i=1$. However, the subleading corrections are different. We next focus on the small $s$ behavior of the characteristic function of the occupation probability of the origin of an active random walker provided in equation~(\ref{asym0n}). The terms appearing at $O(s^i)$ where $i$ is a non-negative integer,  in the small $s$ expansion of $\tilde F(0,s)$ do not contribute to the asymptotic limit of $F(0,t)$. It is clear from the expressions of the coefficients that the leading term that contributes to the first passage probability density in the asymptotic limit ($i={3}/{2}$) itself is not simply derived from a Brownian motion with a modified diffusion constant. i.e., $\phi_\frac{3}{2}(\mathcal{D}_{1d}) \ne \tilde {\phi}_\frac{3}{2}$, where $\tilde{\phi}_{\frac{3}{2}}$ and ${\phi_{\frac{3}{2}}}$ are the coefficients appearing at $O(s^{\frac{3}{2}})$ in~equations~(\ref{asym0n}) and~(\ref{f0ssrw_series}) respectively.

We next study the small and large time limits of the first return probability density by performing a term by term Laplace inversion of equation~(\ref{asym0n}). This yields the limiting forms
\begin{eqnarray}
\label{eqa11}
F(0,t)&\xrightarrow[t \rightarrow \infty]{}& \tilde{\phi}_\frac{3}{2}\frac{1}{t^{\frac{3}{2}}}+{ \tilde{\phi}_\frac{5}{2}}\frac{1}{t^{\frac{5}{2}}}+...~,\nonumber\\
F(0,t)&\xrightarrow[t \rightarrow 0]{}& \tilde{\rho}_1 t + \tilde{\rho}_2 {t^2} + \tilde{\rho}_3 {t^3}+...~.
\end{eqnarray}
\begin{figure}[t!]
\centering
\includegraphics[width=0.95\linewidth]{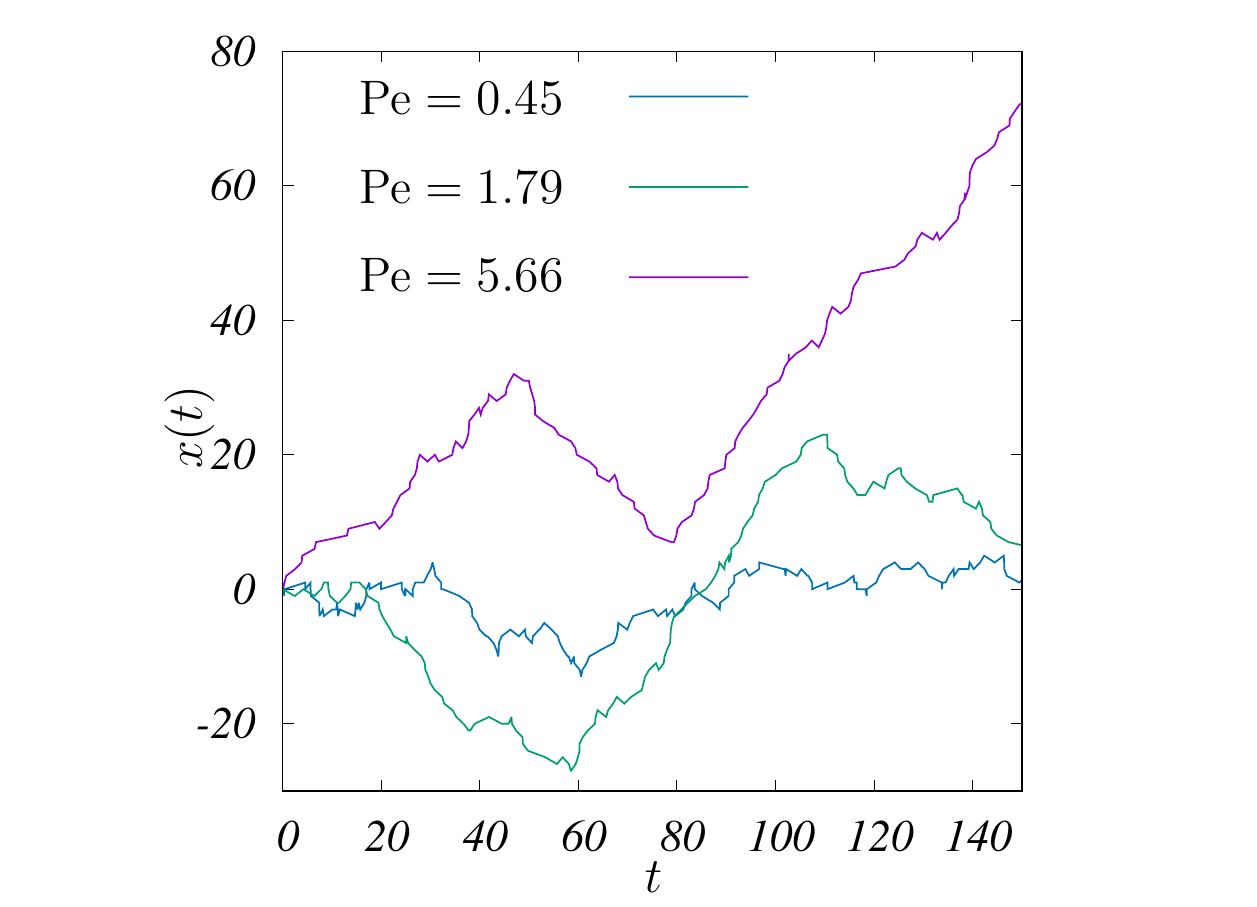}
\caption{Typical trajectories of a RTP in one dimension for different choices of P\'eclet number Pe. For large values of the P\'eclet number, RTP motion is composed of a series of long sojourns with less frequent tumbles.}\label{fig_trajectory}
\end{figure}
The leading order term in the asymptotic limit of the first return probability density has the explicit form
\begin{equation} 
F(0,t)\xrightarrow[t \rightarrow \infty]{}\frac{{\mathcal{D}}_{1d}}{D_{1d}} \sqrt{\frac{ 1}{4\pi {\mathcal{D}}_{1d} }}\frac{1}{t^{\frac{3}{2}}},
\label{eqa1asym}
\end{equation}
where ${\mathcal{D}}_{1d}$ is the effective diffusion constant defined in equation~(\ref{d1dn}) and ${{D}}_{1d}$ is the intrinsic diffusion constant associated with the particle motion. Thus, the probability of the first return to the origin is also governed by the ratio of the diffusion constants ${\mathcal{D}}_{1d}/{D_{1d}}$ which is not predicted by an effective Brownian approximation. Next, we rewrite~equation~(\ref{eqa1asym}) as
\begin{eqnarray} 
F(0,t)&\xrightarrow[t \rightarrow \infty]&{}\sqrt{\frac{{\mathcal{D}}_{1d}}{D_{1d}}} \sqrt{\frac{ 1}{4\pi {{D}}_{1d} }}\frac{1}{t^{\frac{3}{2}}}=\sqrt{1+2 {\text{Pe}}^2} \sqrt{\frac{ 1}{4\pi {{D}}_{1d} }}\frac{1}{t^{\frac{3}{2}}},
\label{eqa1asym2}
\end{eqnarray}
where we define the P\'eclet number Pe, as
\begin{equation}
\text{Pe}=\frac{\epsilon}{\sqrt{D_{1d} \gamma}}.
\label{pecle}
\end{equation}
The P\'eclet number compares the distance travelled between two consecutive tumbles due to drift $\epsilon/\gamma$ to the distance travelled due to diffusive dynamics $\sqrt{D_{1d}/\gamma}$~\cite{kourbane2018exact}. 
When the P\'eclet number $\text{Pe}=0$, we recover the first return probability of an ordinary Brownian motion with diffusion constant $D_{1d}$. For non zero value of the P\'eclet number, we see an enhancement in the first return probability at large times. When the P\'eclet number is large, the motion consists of a sequence of long runs with less frequent changes in the direction.  This increases the probability of the first return at late times. However, when the P\'eclet number is small, the motion has a greater contribution from diffusion, and the first return mostly happens at early times (refer to figure~\ref{fig_asym_srw_arw}).

\begin{figure} [t!]
\hspace{-1 cm}
 \includegraphics[width=1.15\linewidth]{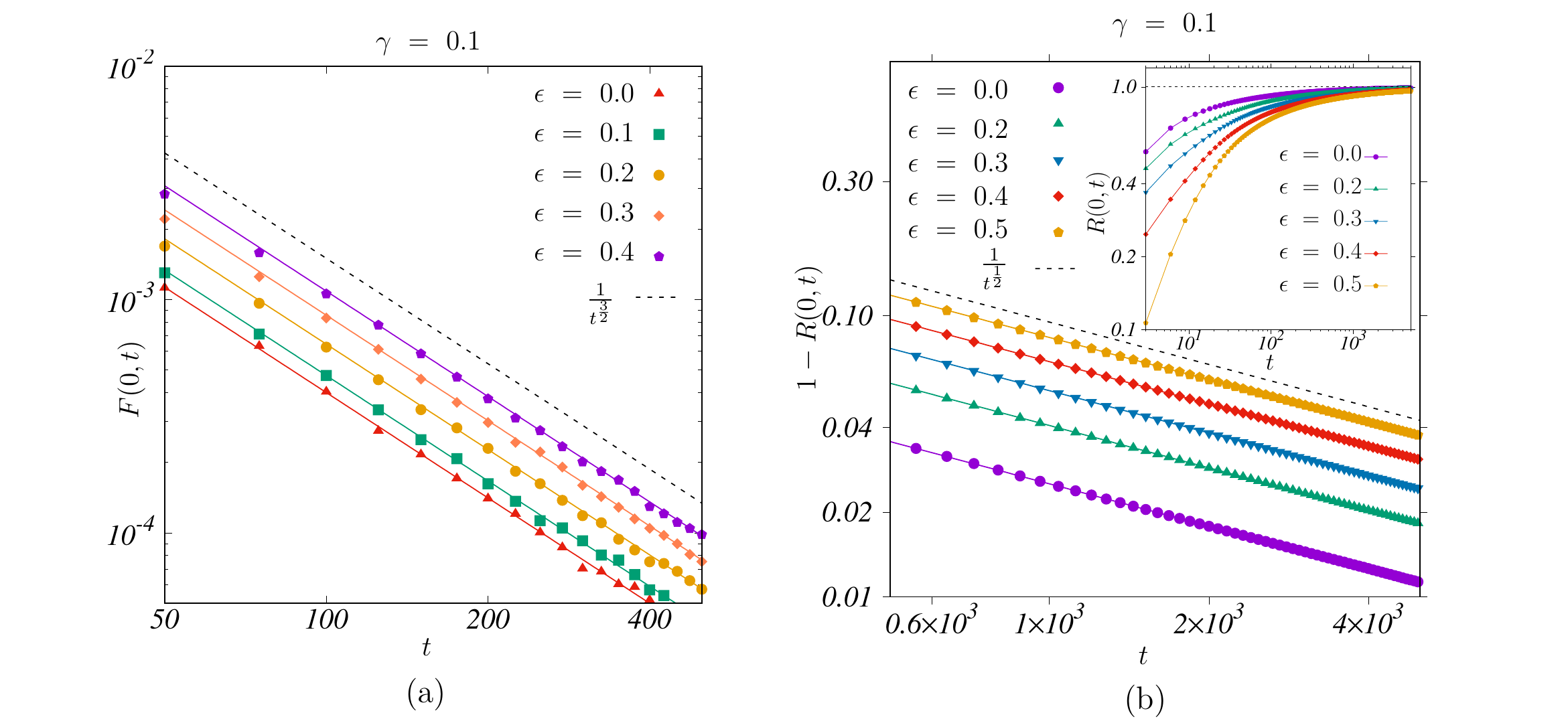}
\caption{(a) Continuous time kinetic Monte Carlo simulation results (points) for the asymptotic limit of the first return probability to the origin $ F(0,t)$, of a RTP on a one dimensional infinite lattice plotted against the theoretical results (solid curves) for the asymptotic limit of $ F(0,t)$ provided in equation~(\ref{eqa1asym}) for different values of $\epsilon$.~(b) Continuous time kinetic Monte Carlo simulation results (points) for the asymptotic limit of the cumulative first return probability $R(0,t)$, for a RTP on a one dimensional infinite lattice plotted against the theoretical results (solid curves) in equation~(\ref{asym0}) for different values of $\epsilon$. Inset:~The inset demonstrates that the probability of ever returning to the origin is $1$ and the walk is recurrent. The fixed parameter values used are $\gamma=0.1$ and $D_{1d}=0.5$.}
\label{fig:f0t_r0t}
\end{figure}

In figure~\ref{fig:f0t_r0t}(a), we compare the long time simulation results for the first return probability to the origin $F(0,t)$, with the leading order term in the asymptotic expansion of $F(0,t)$, provided in equation~(\ref{eqa1asym}).
We also examine the asymptotic limit of the cumulative first return probability $R(0,t)$, of an active random walker in one dimension. The leading order term in the asymptotic expansion of $R(0,t)$ can be found as
\begin{equation}
\label{asym0}
R(0,t)\xrightarrow[t \rightarrow \infty]{} 1-\frac{{\mathcal{D}}_{1d}}{D_{1d}} \sqrt{\frac{ 1}{\pi {{\mathcal{D}}_{1d}}}}\frac{1}{t^{\frac{1}{2}}}.
\end{equation}
 From equation~(\ref{asym0}), it is clear that an active walk in one dimension is recurrent for any non-zero $\gamma$. That is, $R(0,\infty)=1$ and the walk is recurrent. We display a plot comparing the simulation results for the cumulative first return probability density $R(0,t)$ and the analytic expression provided in~equation~(\ref{asym0}), for fixed parameter values $\gamma=0.1$ and $D_{1d}=0.5$ in figure~\ref{fig:f0t_r0t}(b). The theoretical predictions and the simulation results are in excellent accord. 
\subsection{For \texorpdfstring{$x \ne 0$}{Lg}}
We next analyze the probability of the first passage of an active random walker to an arbitrary site $x$, different from the origin.
We obtain the exact expression for the Laplace transform of the probability of the first passage to any arbitrary site $x \ne 0$, by substituting the expressions for the Laplace transforms of the occupation probabilities in equation~(\ref{reca0}). Since this expression is quite long, we do not quote it here. Nevertheless, the small $s$ behavior of $\tilde F(x,s)$ can be obtained as
\begin{eqnarray} 
\label{series_fxs}
\hspace{-1.0 cm}
\tilde F(x,s)&\xrightarrow[s \rightarrow 0]{}&\exp\left(-\left| x\right| \sqrt{\frac{s}{{\mathcal{D}}_{1d}}} \right) \left [ \tilde{\xi}_1s^0-2\tilde{\xi}_\frac{3}{2}\sqrt{\pi}s^{\frac{1}{2}}+\tilde{\xi}_2 s^1+...\right ].
\end{eqnarray}
The set of coefficients $\{\tilde{\xi}_i\}$ appearing in the above equations are functions of the rates $\gamma,~\epsilon$ and $D$ and the variable $x$. We provide explicit forms of the first few coefficients $\{\tilde{\xi}_{i}\}$ in table~\ref{table_arw} of appendix~\ref{appendix_a}. The main aim of this study is to analyze the behavior of the first passage probability density of an active random walk, as well as the nature of its deviation from results for a symmetric random walk with an effective diffusion constant at large times. The expression in equation~(\ref{series_fxs}) is to be compared with the corresponding expression for a symmetric random walk. Therefore, in order to study the first passage statistics of active random walks, it is extremely useful to have the first passage probability density in Laplace space. 
\begin{figure} [t!]
\hspace{-1 cm}
 \includegraphics[width=1.15\linewidth]{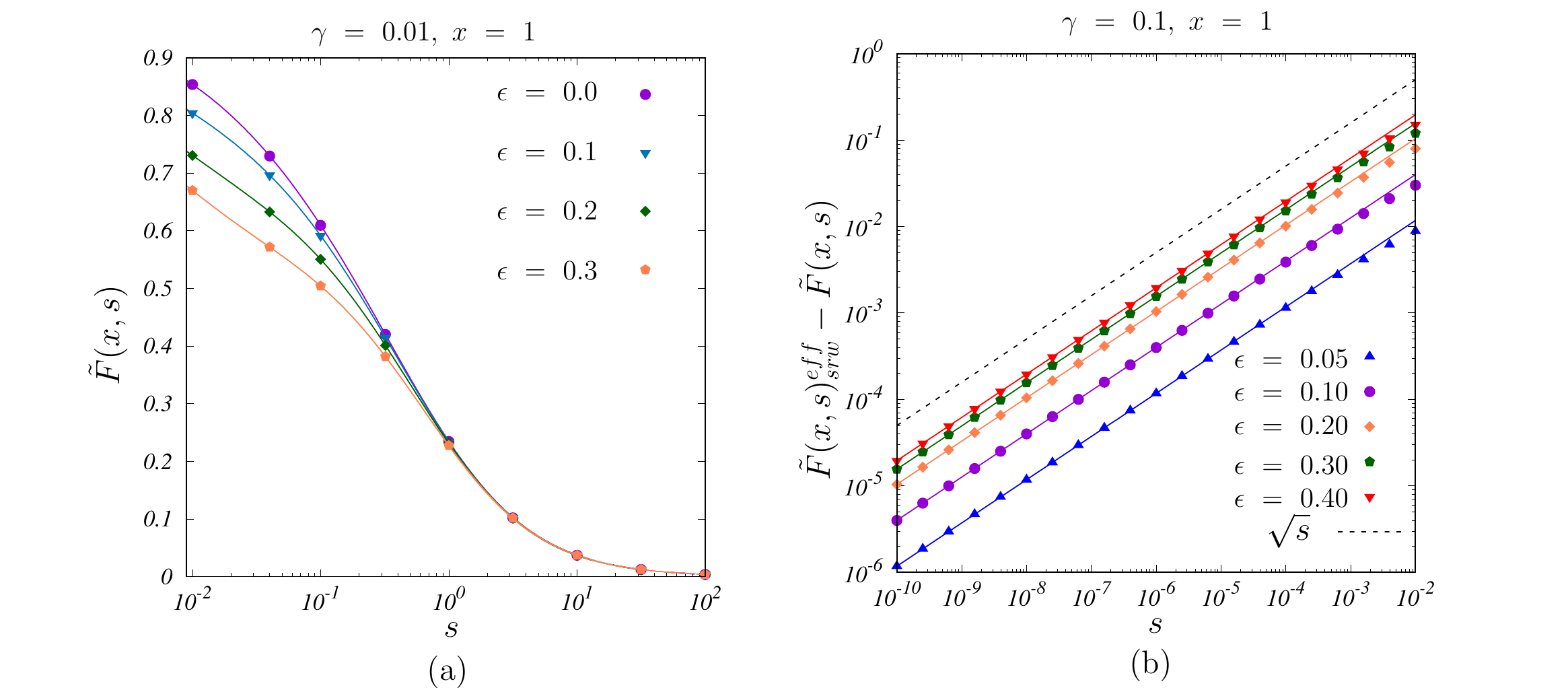} 
\caption{(a) The Laplace transform of the first passage probability distribution of an active random walker $\tilde F(x,s)$, obtained from kinetic Monte Carlo simulations (points) plotted as a function of $s$ for the lattice site $x=1$, for different values of $\epsilon$. The simulation data is averaged over $10^7$ realizations. The solid curves for non-zero $\epsilon$ are obtained by substituting the expressions for the characteristic functions for the occupation probabilities in equation~(\ref{reca0}). The solid curve for $\epsilon=0.0$ which in turn corresponds to the limit of a symmetric random walker is obtained from the exact analytic expression provided in equation~(\ref{fxssrw}). The fixed parameter values used are $\gamma=0.01$ and $D_{1d}=0.3$.~(b) The difference between the Laplace transforms of the first passage probability distribution of a symmetric random walker with an effective diffusion ${\tilde F(x,s)}^{eff}_{srw}$ obtained by substituting $D_{1d}=\mathcal{D}_{1d}$ in equation~(\ref{fxssrw}) and an active random walker $\tilde F(x,s)$, provided in equation~(\ref{series_fxs}) plotted (points) as a function of $s$ for the lattice site $x=1$, for different values of $\epsilon$. The solid curves are obtained from the exact analytic expressions provided in equation~(\ref{diff_fun_fxs_1d}). We observe that both the distributions are different at the leading order $\left [O(\sqrt{s}) \right]$. The fixed parameter values used are $\gamma=0.1$ and $D_{1d}=0.45$.}
\label{fig:fxs_fxs_conv}
\end{figure}
\subsubsection{Limiting cases}

For a biased random walker ($brw$), we obtain the characteristic function of the probability of the first passage to an arbitrary lattice site $x \ne 0$ as
\begin{equation}
\label{fxsbrw}
 {\tilde F(x,s)}_{brw}=
\frac{{\left(s+2 D_{1d}-\sqrt{s(s+4 D_{1d})+4 \epsilon^2}\right)}^{\left| x\right|}}{2^{\left| x\right|+1}}{\left[\frac{1}{{(D_{1d}- \epsilon)}^{\left| x\right|}}+\frac{1}{{(D_{1d}+ \epsilon)}^{\left| x\right|}}\right]}.
\end{equation} 
This expression can be obtained from the corresponding expression for an active random walker by substituting the flip rate $\gamma=0$. We obtain the characteristic function of the first passage probability distribution of a symmetric random walker ($srw$) by substituting the bias rate $\epsilon=0$ in the above equation as
\begin{equation}
\label{fxssrw}
 {\tilde F(x,s)}_{srw}=\frac{\left(s+2 D_{1d}-\sqrt{s (s+4 D_{1d})}\right)^{\left| x\right| }}{(2 D_{1d})^{\left| x\right| } }.
\end{equation} 
 In figure~\ref{fig:fxs_fxs_conv}(a), we display the convergence of the characteristic function of the first passage time density of an active random walker $\tilde F(x,s)$, to that of a symmetric random walker ${\tilde F(x,s)}_{srw}$ as the bias rate $\epsilon \rightarrow 0$.
The first passage probability densities for a passive random walker in the  time domain are obtained after performing a Laplace inversion of the above equations. This results in the following expressions~\cite{khantha1983first,balakrishnan1983some,jose2021first},
\begin{small}
\begin{equation}
\label{fxtbrw}
{F(x,t)}_{brw}=\frac{e^{-2D_{1d}t}}{2}\frac{{|x|}}{t}I_{\left |x  \right |}\left(2 \sqrt{\eta}~t \right)\left [\sqrt{{\left(\frac{(D_{1d}+ \epsilon)}{(D_{1d}- \epsilon)}\right)}^{\left| x\right|}}+\sqrt{{\left(\frac{(D_{1d}- \epsilon)}{(D_{1d}+ \epsilon)}\right)}^{\left| x\right|}} ~\right ],
\end{equation}
\end{small}
\begin{equation}
\label{fxtsrw}
{F(x,t)}_{srw}=e^{-2 D_{1d} t}\frac{{|x|}}{t} I_{\left| x\right|}\left(2 {{D_{1d}}} t\right).
\end{equation}

Since we are interested in studying the deviation of an active random walk from a symmetric random walk in the long time limit in terms of the first passage probabilities, we perform a series expansion of the expression provided in equation~(\ref{fxssrw}) around small $s$ to obtain
\begin{eqnarray} 
\label{series_fxs_srw}
\hspace{-1.0 cm}
\tilde F(x,s)_{srw}&\xrightarrow[s \rightarrow 0]{}&\exp\left(-\left| x\right| \sqrt{\frac{s}{{{D}}_{1d}}} \right) \left [ {\xi_1}s^0+\frac{4}{3}{\xi_\frac{5}{2}}\sqrt{\pi} s^{\frac{3}{2}}+...\right ].
\end{eqnarray}
We next analyze the difference between the first passage probabilities of an active random walk and a symmetric random walk with an effective diffusion constant in the asymptotic limit by examining the corresponding distributions in Laplace space. The expression for the characteristic function of the first passage distribution of a symmetric random walk with an effective diffusion denoted as ${ \tilde F(x,s)}^{eff}_{srw}$, is obtained simply by substituting $D_{1d}={\mathcal{D}}_{1d}$ in equation~(\ref{fxssrw}). We display a plot for the difference between the characteristic functions of a symmetric random walk with a modified diffusion constant${ \tilde F(x,s)}^{eff}_{srw}$, and an active random walk ${\tilde F(x,s)}$, in the small $s$ regime for the lattice site $x=1$, in figure~\ref{fig:fxs_fxs_conv}(b). We notice that for small $s$, the difference grows at the leading order $O(\sqrt{s})$. This is because, the actual behavior of $\tilde F(x,s)$ in the small $s$ limit is also determined by the second term in equation~(\ref{series_fxs}), which also appears at $O(\sqrt{s})$. The exact behavior can be calculated analytically by using the expressions for the associated Laplace transforms and we obtain
\begin{equation}
\label{diff_fun_fxs_1d}
 { \tilde F(x,s)}^{eff}_{srw}-\tilde F(x,s) \xrightarrow[s \rightarrow 0]{}\begin{cases}
-2\sqrt{\pi}\left[{\phi_\frac{3}{2}(\mathcal{D}_{1d})}-\tilde{\phi}_\frac{3}{2}\right] \sqrt{s},& x= 0, \\\\
-2\sqrt{\pi}\tilde{\xi}_\frac{3}{2} \sqrt{s}, & x\ne 0.
\end{cases}
\end{equation}
\begin{figure} [t!]
\hspace{-1 cm}
 \includegraphics[width=1.15\linewidth]{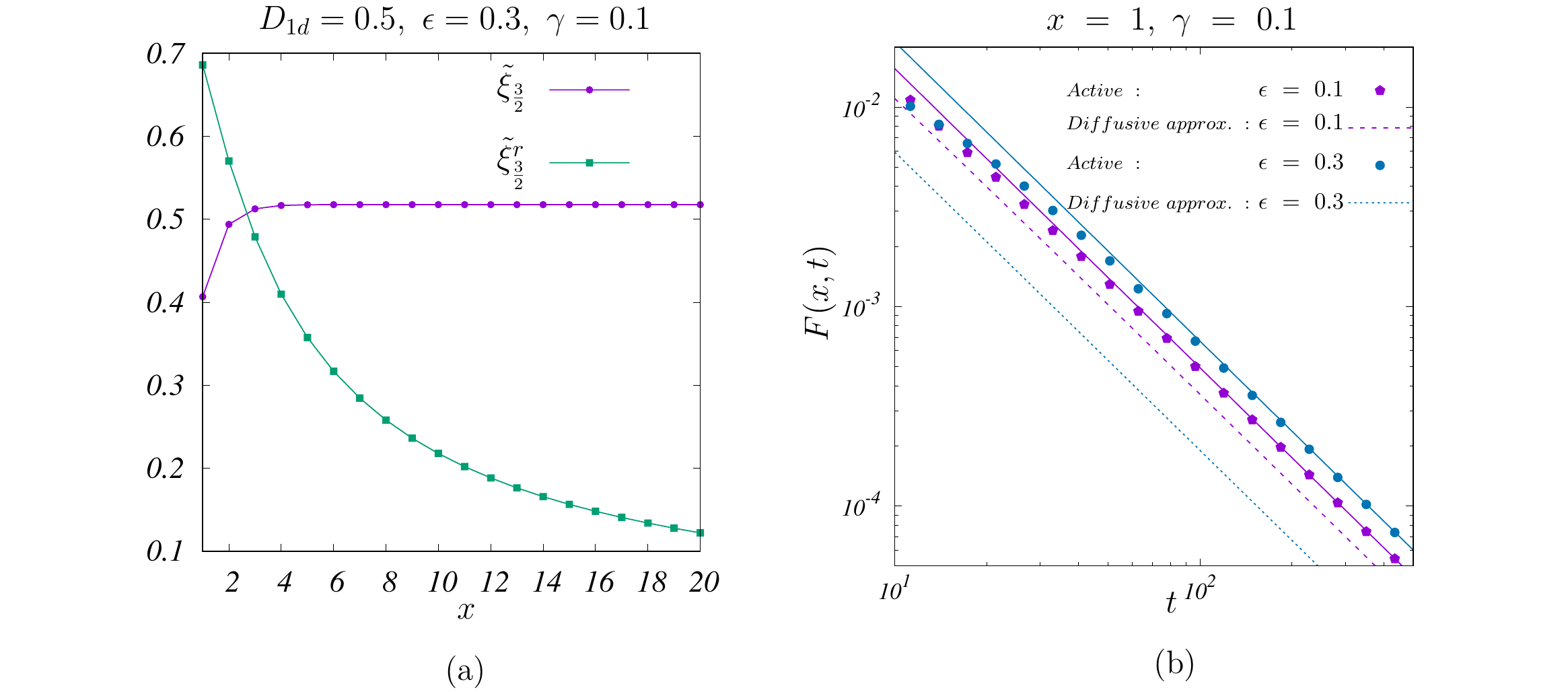}  
\caption{(a) The coefficient $\tilde{\xi}_\frac{3}{2}$ (provided in table~\ref{table_arw} of appendix~\ref{appendix_a}) plotted as a function of $x$ for fixed parameter values $D_{1d}=0.5,~\epsilon=0.3,~\gamma=0.1$. We notice that $\tilde{\xi}_\frac{3}{2}$ reduces to a constant for lattice sites away from the origin. The coefficient $\tilde{\xi}_\frac{3}{2}$ is symmetric in the variable $x$.  We also plot $\tilde{\xi}^r_\frac{3}{2}=\tilde{\xi}_\frac{3}{2}/( \frac{\left| x\right|}{\sqrt{4 \pi \mathcal{D}}_{1d}}+\tilde{\xi}_\frac{3}{2})$ which gives a measure of the relative magnitude of the correction due to activity. This correction is extremely large for the lattice sites close enough to the origin.~(b)~The first passage time density of an active random walker $ F(x,t)$ in one dimension, obtained from kinetic Monte Carlo simulations (points) plotted as a function of time for the lattice site $x=1$, for two different values of $\epsilon$. The solid curves correspond to the theoretical result in equation~(\ref{series_fxt_actual}). The dashed curves correspond to an ordinary one dimensional Brownian motion with an effective diffusion constant~$\mathcal{D}_{1d}$. The fixed parameter values used are $\gamma=0.1$ and $D_{1d}=0.40$.}
\label{fig:xi_fxt}
\end{figure}
The explicit forms of the coefficients ${\phi_{\frac{3}{2}}},~\tilde{\phi}_{\frac{3}{2}}$ and $\tilde{\xi}_{\frac{3}{2}}$ are provided in tables~\ref{table_arw} and~\ref{table_brw} of appendix~\ref{appendix_a}. After Laplace inversion, we obtain the limiting behaviors of the probability of the first passage of an active random walker to  a lattice site $x$, as
\begin{equation} 
\label{series_fxt_actual}
F(x,t)\xrightarrow[t \rightarrow \infty]{}
\begin{cases}
\tilde{\phi}_\frac{3}{2}\frac{1}{t^{\frac{3}{2}}},& x= 0, \\\\
\left (\frac{\left| x\right|}{\sqrt{4 \pi \mathcal{D}_{1d}}}+\tilde{\xi}_\frac{3}{2} \right)\frac{1}{t^{\frac{3}{2}}}, & x\ne 0.
\end{cases}
\end{equation}
In the above equation, the coefficient $\tilde{\xi}_\frac{3}{2}$ is a function of the variable $x$. For $x \ne 0$, the first term in the RHS of~equation~(\ref{series_fxt_actual}) can be thought of as coming from an ordinary Brownian motion with the modified diffusion constant $\mathcal D_{1d}$. However, the correction $\tilde{\xi}_\frac{3}{2}$ is indicative of non trivial signatures of activity at large times. This correction is extremely large for the lattice sites close enough to the origin~[refer to figure~\ref{fig:xi_fxt}(a)]. For the lattice sites very far from the origin ($x \rightarrow \infty$), the coefficient~$\tilde{\xi}_\frac{3}{2}$ reduces to the constant,
\begin{equation}
\lim_{x \rightarrow \infty}
\tilde{\xi}_\frac{3}{2}=\tilde{\xi}^c_\frac{3}{2}=\frac{\sqrt{\frac{1}{D_{1d}}+\frac{2}{\gamma }} \epsilon ^2}{\sqrt{4 \pi } \gamma  \mathcal{D}_{1d}}.
\label{xi_c}
\end{equation}
\begin{figure}[t!]
\centering
\includegraphics[width=0.95\linewidth]{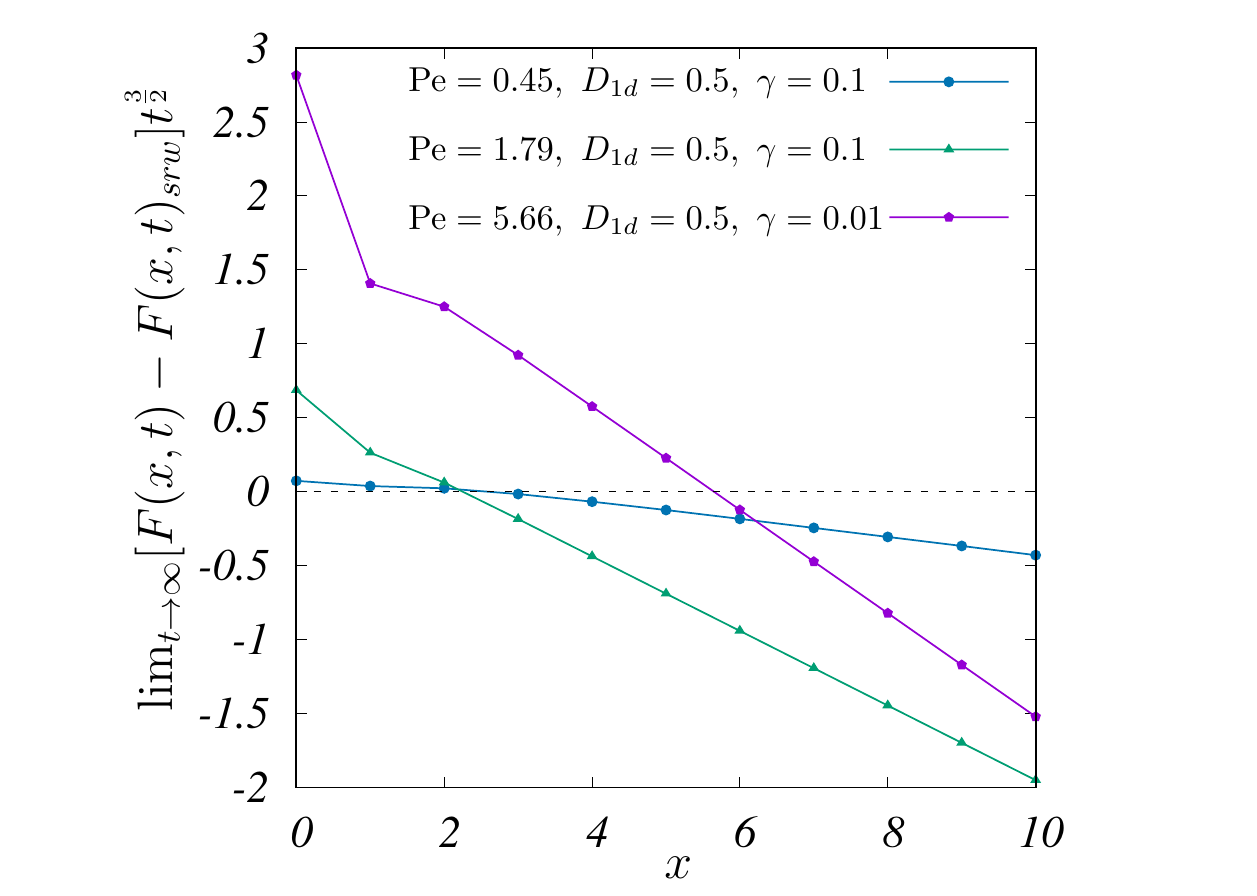}
\caption{The difference between the first passage probabilities of an active random walker, and a symmetric random walker with diffusion constant $D_{1d}$ provided in equation~(\ref{fxt_dif}) plotted as a function of $x$ for different values of P\'eclet number Pe. When the magnitude of the difference function $\lim_{t \rightarrow \infty}[F(x,t)-F(x,t)_{srw}]t^{\frac{3}{2}}$ is positive, there is an enhancement in the first passage probabilities at late times due to activity.}\label{fig_asym_srw_arw}
\end{figure}

To construct the limiting form in~equation~(\ref{series_fxt_actual}), we take the $t \rightarrow \infty$ limit keeping $x$ fixed. 
Another scaling limit of interest is the $x,~t \rightarrow \infty$ limit, with $x/t^{\frac{1}{2}}$ held fixed. In the latter case, the ﬁrst-passage time density reduces to that of an ordinary
Brownian motion with a modified diffusion constant $\mathcal D_{1d}$ and the results of~\cite{malakar2018steady} are reproduced. Since we are also interested in the probability of the first return to the origin as well as the probability of the first passage to lattice sites close enough to the origin, the results in~equation~(\ref{series_fxt_actual}) are particularly useful. The expression provided in~equation~(\ref{series_fxt_actual}) for non zero $x$ is very similar to the expression for the survival probability of a RTP derived in~\cite{le2019noncrossing} in one dimension. However in that case, a continuous space model for a purely active process (zero diffusion) is considered. The analogue of the coefficient $\tilde{\xi}_\frac{3}{2}$ in the zero diffusive limit can be mapped onto the Milne extrapolation length known in nuclear physics. It has been shown that this coefficient is independent of $x$ in the zero diffusive case. However, any finite value of $D_{1d}$ introduces non trivial dependence on the space variable $x$. A discussion of this result is provided in appendix~\ref{appendix_b} where we describe the continuum version of the discrete space model of RTP.  

As a result, we infer that at large times, the first passage probability density of an active random walk does not perfectly match that of a Brownian particle with an effective diffusion constant.
In the case of occupation probabilities, on the other hand, the leading order behavior of an active random walk is correctly reproduced by the case of a symmetric random walk with an effective diffusion constant over large time scales. This occurs since the first passage distribution is given by a {\it ratio} of the occupation probabilities as in equation~(\ref{reca0}), which gives rise to deviations from the symmetric random walk case, even at the leading order.
We expect the influence of activity in the first passage probability to endure in a nontrivial fashion even at long intervals since the first passage is a one time event and is subjected to less averaging. In figure~\ref{fig:xi_fxt}(b), we compare our theoretical prediction for the asymptotic
behavior of the first passage probability density of an active random walk in one
dimension provided in equation~(\ref{series_fxt_actual}) with simulation results for the lattice site $x = 1$. We
notice that effective diffusive approximation [without the second term in the RHS of
the second expression provided in equation~(\ref{series_fxt_actual})] does not capture the right behavior of the
first passage probabilities of an active random walk at large times. We also verify this departure from the effective Brownian picture for first passage probabilities in continuous space by performing Monte Carlo simulations in continuous space (refer to appendix~\ref{appendix_b}).

In our analysis of the first return probability to the origin, we have observed an increase in the probability of the first return to the origin at large times owing to activity.
As a result, it would also be interesting to investigate how activity influences the  first passage to arbitrary lattice positions. 
From equation~(\ref{series_fxt_actual}), it is possible to write 
\begin{equation}
  \lim_{t \rightarrow \infty }  [F(x,t)]-F(x,t)_{srw}]t^{\frac{3}{2}}= \left (\frac{\left| x\right|}{\sqrt{4 \pi \mathcal{D}_{1d}}}+\tilde{\xi}_\frac{3}{2} \right)-\frac{\left| x\right|}{\sqrt{4 \pi {D}_{1d}}},
  \label{fxt_dif}
\end{equation}
where $F(x,t)$ is the first passage probability density of an active random walker and $F(x,t)_{srw}$ is the first passage probability density of a symmetric random walker with diffusion constant $D_{1d}$.
We display a plot for this difference function at large times in figure~\ref{fig_asym_srw_arw}. We see that activity increases the likelihood of the first passage to lattice sites close enough to the origin at large time. However, for lattice sites far enough from the origin, activity reduces the first passage probabilities at large time. 
A quantitative estimate of the scale of $x$ at which this crossover happens ($x_{cross}$) can be obtained by replacing $\tilde{\xi}_\frac{3}{2}$ by $\tilde{\xi}^c_\frac{3}{2}$ in the RHS of above equation, and equating the RHS to zero. Thus we obtain
\begin{equation}
    x_{cross} \approx \sqrt{1+2 \frac{D_{1d}}{\gamma}}\frac{\text{Pe}^2}{[(1+2 \text{Pe}^2)-\sqrt{1+2 \text{Pe}^2}]},
    \label{x_cross}
\end{equation}
where the P\'eclet number is defined in equation~(\ref{pecle}). The value of $x_{cross}$ is not a function of the P\'eclet number alone, but also depends on the diffusive and flipping rates as is evident from the above equation. 

We utilize these insights in the next section where we study the first passage probabilities of a two dimensional active random walk. In this case, it is difficult to derive the correction to the leading order result in closed-form for arbitrary lattice sites. However, using the exact expressions in Fourier-Laplace space, we show numerically, that the leading order behavior of the first passage density is not accurately captured by a symmetric random walk.

\section{Occupation probabilities of active random walks in two dimensions}
\label{subsec:occ_prob_2d}
We next consider the motion of an active random walker starting from the origin at time $t=0$, on a two dimensional infinite square lattice. We assume symmetric initial conditions where the particle has equal initial probabilities (${1}/{4}$ each) to be in any of the four possible internal states $0,~1,~2$ or $3$ at time $t=0$. The particle is weakly biased along the $+x$, $+y$, $-x$ or $-y$ direction if it is in state $0,~1,~2$ or $3$ respectively. The evolution equation for the probability of occupation of a lattice site $(x,y)$, by an active particle in the internal bias direction $m$ denoted as $P_m \equiv P_m (x,y,t)$ is provided in equation~(\ref{sq0}). The total probability to occupy a site $(x,y)$ at time $t$, is given as $P(x,y,t)=\sum_{m=0}^3 P_m(x,y,t)$.
We begin by studying the Fourier-Laplace transform of the occupation probability $P(x,y,t)$of a lattice site $(x,y)$, in two dimensions defined as $\tilde P (k_x,k_y,s)=\sum_{x=-\infty}^{\infty}\sum_{y=-\infty}^{\infty}\int_0^{\infty}dte^{i(k_x x+k_y y)-s t}P(x,y,t)$. This can be obtained by taking a Fourier-Laplace transform of equation~(\ref{sq0}). After simplification, we obtain as in \cite{PhysRevE.105.064103}
\begin{eqnarray}
\label{pks2dn}
\hspace{-2.5 cm}
\tilde P(k_x,k_y,s)=\biggl[ 4 f^2+(s+\gamma ) (s+2 \gamma )-2 f (2 s+3 \gamma )+\epsilon ^2(2-g) \biggr]/\nonumber\\\hspace{-2.5 cm}\biggl[2\epsilon ^2 (2-g) (-2 f+s+\gamma ) +\frac{
   \epsilon ^4 h^2}{(-2 f+s+\gamma )}+(2 f-s)(2 f-s-2 \gamma ) (-2 f+s+\gamma )\biggr],\nonumber\\
\end{eqnarray}
where 
\begin{equation}
 g \equiv g(k_x,k_y)=\cos (2 k_x) +\cos (2 k_y),   
\end{equation}
\begin{equation}
f \equiv f(k_x,k_y)=D_{2d}\left[-2+g\left(\frac{k_x}{2},\frac{k_y}{2}\right)\right],    
\end{equation}
and
\begin{equation}
\label{hx}
h \equiv h(k_x,k_y)= 4 \sin k_x \sin k_y.  
\end{equation}
We need to perform the integration,
\begin{equation} 
\tilde P(x,y,s)=\frac{1}{4 \pi^2}\int_{-\pi}^{\pi} \int_{-\pi}^{\pi} dk_x dk_ye^{-i(k_x x+k_y y)} \tilde P(k_x,k_y,s), 
\label{eq:fourier_inv}
\end{equation}
to obtain the Laplace transform of the occupation probability of a lattice site $(x,y)$. Unfortunately, it is hard to perform this integration exactly and obtain a closed-form expression for $\tilde P(x,y,s)$. However, it is possible to extract the asymptotic behavior of the occupation probability $P(x,y,t)$ from the exact expression for the characteristic function provided in equation~(\ref{pks2dn}).
\begin{figure} [t!]
\hspace{-0.4 cm}
 \includegraphics[width=1.10\linewidth]{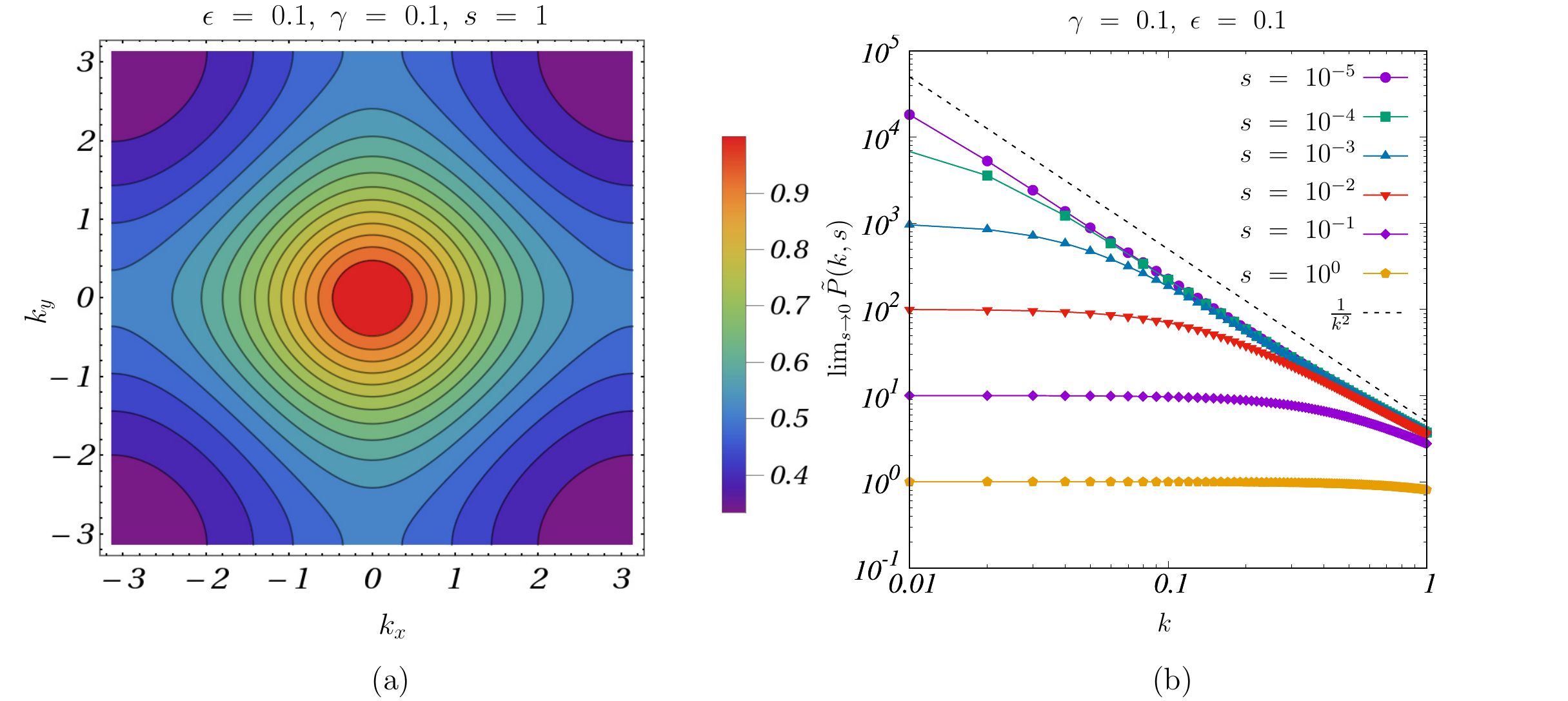} 
\caption{(a) The Fourier-Laplace transform of the occupation probability of a lattice site in two dimensions $\tilde P(k_x,k_y,s)$, provided in equation~(\ref{pks2dn}), plotted for fixed parameter values $\gamma=0.1$, $\epsilon=0.1$ and $D_{2d}=0.25$. Here, we have set $s=1$. We observe that $\tilde P(k_x,k_y,s)$ is rotationally invariant in the limit $k_x\rightarrow0$,~$k_y\rightarrow0$. i.e, the system can be described by a single $k=\sqrt{{k_x}^2+{k_y}^2}$ when $k_x$ and $k_y$ have small values. The symmetry of the lattice (four-fold symmetry) emerges at intermediate values of $k_x$ and $k_y$.~(b) The Fourier-Laplace transform of the occupation probability of a lattice site in two dimensions of a RTP in the small $k$ limit $\tilde P(k,s)$, provided in equation~(\ref{series_k1}) plotted as a function of $k$ for small $s$ values. We observe that $\lim_{s\rightarrow 0}\tilde P(k,s)$  scales as $\frac{1}{s}$ for small $k$ values and then falls as $\frac{1}{\mathcal{D}_{2d}k^2}$. The fixed parameter values used in the plot are $\gamma=0.1$,~$\epsilon=0.1$ and $D_{2d}=0.25$.}
\label{fig:free_energy_pks}
\end{figure}
In figure~\ref{fig:free_energy_pks}(a), we display a typical plot of $\tilde P(k_x,k_y,s)$ as a function of $k_x$ and $k_y$ for fixed $s$. The function $\tilde P(k_x,k_y,s)$ is radially symmetric in the limit $k_x\rightarrow0$,~$k_y\rightarrow0$ and we obtain
$
\tilde P(k_x,k_y,s)\xrightarrow[k_x\rightarrow 0,~k_y\rightarrow 0]{}\tilde P(k,s),
$
where $k=\sqrt{k_x^2+k_y^2}$. In the limit of small $k_x,~k_y$, we obtain an explicit form for this radial function as
\begin{equation}
\label{series_k1}
\tilde P(k,s)=\frac{  (s+\gamma ) (s+2 \gamma )+D_{2d}k^2 \left(3 s+5 \gamma\right) +2 k^2 \epsilon ^2}{ 2 D_{2d} k^2 \left(2 s^2+4 s \gamma +\gamma ^2\right)+(s+\gamma ) \left(s^2+2 s \gamma +4 k^2 \epsilon ^2\right)}.
\end{equation}
For any non-zero $\gamma$, $\lim_{s\rightarrow 0}\tilde P(k,s)$  scales as $\frac{1}{s}$ for small $k$ values and then it decays as $\frac{1}{\mathcal{D}_{2d}k^2}$, where $\mathcal{D}_{2d}$ is the effective diffusion constant~\cite{howse2007self,lindner2008diffusion,solon2015active,aragones2018diffusion,basu2018active,PhysRevE.105.064103} in two dimensions defined as
\begin{equation}
 \mathcal{D}_{2d}=   {D}_{2d}+\frac{2 \epsilon^2}{\gamma}.
 \label{d2d}
\end{equation} 
In figure~\ref{fig:free_energy_pks}(b), we display a plot of $\tilde P(k,s)$ with the two regimes in the limit of small $s,~k$. 
The leading order term (for non-zero $\gamma$) in the small $s$ expansion of the Laplace transform of the occupation probability of the origin $\tilde P (0,s)$, can be found by performing an inverse Fourier transform of equation~(\ref{series_k1}) for $x=0,y=0$. This yields
\begin{equation} 
\label{ser2d}
\tilde P (0,s)\xrightarrow[s \rightarrow 0]{}\--\frac{{\tilde{\lambda}_1}}{4\pi}\log s+O(s^0)~,
\end{equation}
where $\tilde{\lambda}_1= \frac{1}{{\mathcal{D}_{2d}}}$.
The Laplace inversion of the function in equation~(\ref{ser2d}) should give the leading behavior of the occupation probability in the long time limit. The constant correction to $\tilde P (0,s)$ for small $s$ does not affect the large time behavior of $P(0,t)$. We thus obtain
\begin{equation} 
\label{eqa12d}
\hspace{-0.6 cm}
P(0,t)\xrightarrow[t \rightarrow \infty]{}\ L^{-1}[- \frac{{\tilde{\lambda}_1}}{4\pi}\log s+...]= \frac{1}{4\pi \mathcal{D}_{2d}}\frac{1}{t}+...~.
\end{equation}
This resembles the occupation probability of a two dimensional Brownian motion with an effective diffusion constant $\mathcal{D}_{2d}$. 
We also notice that the effective diffusion constant appears with the same correction $\frac{2 \epsilon^2}{\gamma}$, in one dimension [refer to equation~(\ref{d1dn})] and two dimensions [refer to equation~(\ref{d2d})]. 
Since it is difficult to solve the integrals analytically for any arbitrary site $x\ne 0,~y\ne 0$, we use numerical integration techniques. While studying the first passage statistics in two dimensions, we compute the integral in equation~(\ref{eq:fourier_inv}) using numerical integration for lattice points other than the origin. 
\subsection{Limiting cases}

We first analyze various limiting cases of the characteristic function of an active lattice walk in two dimensions.
Setting $\gamma=0$ in equation~(\ref{pks2dn}) yields the Fourier-Laplace transform of the occupation probability of a biased random walker ($brw$) on a two dimensional infinite square lattice. For a walker biased along the positive $x$ direction, we obtain
\begin{equation}
 {\tilde P(k_x,k_y,s)}_{brw}=\frac{1}{4 D_{2 d}+s-2 D_{2 d} (\cos k_x+\cos k_y)+2 i \epsilon  \sin k_x}. 
 \label{eq:brw_pks}
\end{equation}
Similarly, setting $\epsilon=0$ in equation~(\ref{pks2dn}) yields the Fourier-Laplace transform of the occupation probability of a symmetric random walker ($srw$) on a two dimensional infinite square lattice as
\begin{equation} 
{\tilde P(k_x,k_y,s)}_{srw}=\frac{1}{4 D_{2 d}+s-2 D_{2 d} (\cos k_x+\cos k_y)}.
 \label{eq:srw_pks}
\end{equation}
While it is hard to perform the exact inverse Fourier transform of equation~(\ref{eq:brw_pks}), it is possible to invert the Fourier transform in equation~(\ref{eq:srw_pks}) exactly which corresponds to the limiting case of a symmetric random walker.
The Fourier inversion of equation~(\ref{eq:srw_pks}) yields the Laplace transform of the occupation probability of a symmetric random walker on a two dimensional square lattice. This is given as
\begin{equation}
\label{latticegreenfun}
{\tilde P(x,y,s)}_{srw}=\frac{1}{4 {\pi}^2}\int_{-\pi}^{\pi}\int_{-\pi}^{\pi}e^{-i(k_xx+k_yy)}{\tilde P(k_x,k_y,s)}_{srw}dk_x dk_y.
\end{equation}

This integral is the lattice Green's function for a square lattice which was first derived in closed-form in \cite{katsura1971latticea} and then applied in different random walk contexts in~\cite{hughes1995random,maassarani2000series,jose2021first}. We thus obtain the exact analytic expression for the  characteristic function of the lattice occupation probability of a symmetric random walker on a two dimensional infinite square lattice as
\begin{small}
\begin{eqnarray}
\label{eq:x}
\hspace{-2.5 cm}
{\tilde P(x,y,s)}_{srw}= \,
   _4F_3 \biggl[\frac{1+|x|+|y|}{2},\frac{1+|x|+|y|}{2},\frac{2+|x|+|y|}{2},\frac{2+|x|+|y|}{2};1+|x|,1+|y|,\nonumber\\\hspace{-2.5 cm} 1+|x|+|y|;    {\left (\frac{4 D_{2d}}{s+4 D_{2d}} \right)}^2 \biggr]/ \left[\frac{(s+4 D_{2d})^{1+|x|+|y|}}{D_{2d}^{|x|+|y|}} \frac{|x|! |y|!}{(|x|+|y|)!}\right],
\end{eqnarray}
\end{small}
where $ _4F_3$ is the hypergeometric function.
For diagonal sites $x=y$, we obtain the simplified expression
\begin{equation}
\label{hh}
\hspace{-0.2 cm}
{\tilde P(x,s)}_{srw}=\frac{\Gamma \left(\frac{1}{2}+|x|\right) \, _2F_1\left[\frac{1+2|x|}{2},\frac{1+2|x|}{2};1+2
   |x|;{\left (\frac{4 D_{2d}}{s+4 D_{2d}} \right)}^2 \right]}{ \frac{(s+4 D_{2d})^{1+2 |x|}}{{(2D_{2d})}^{2 |x|}} \sqrt{\pi }~ \Gamma (1+|x|)},
\end{equation}
where ${\tilde P(x,s)}_{srw}={\tilde P(x,x,s)}_{srw}$ and  $\Gamma$ is the gamma function. We obtain the Laplace transform of the occupation probability of the origin ${\tilde P(0,s)}_{srw}$ by setting $x=0$ in the above equation as
\begin{equation} 
\label{eq:y}
{\tilde P(0,s)}_{srw}=\frac{2 K\left[{\left (\frac{4 D_{2d}}{s+4 D_{2d}} \right)}^2\right]}{\pi  (s+4 D_{2d})}.
\end{equation}
In the above expression, $K$ is the elliptic integral of the first kind.
We next invert the Laplace transform in equation~(\ref{eq:x}) to obtain \cite{balakrishnan1983some}
\begin{equation} 
\label{eq:w}
 {P(x,y,t)}_{srw}=e^{-4D_{2d}t} {I_{|x|}}\left({2 D_{2d}t}\right){I_{|y|}}\left({2 D_{2d}t}\right).
\end{equation}
The above equation holds for any arbitrary lattice site $(x,y)$ including the origin. Since we are interested in the long time behavior of the first passage distribution of an active random walk, a closed-form expression for the characteristic functions of a symmetric random walk would be helpful to study the convergence to or deviation from an active random walk. Again, this is possible due to the fundamental renewal relations connecting the occupation probabilities and the first passage probability distributions provided in~equations~(\ref{reca0}) and (\ref{reca}).
\section{First passage statistics of active random walks in two dimensions}
\label{subsec:first_passage_2d}

For an active random walker in two dimensions, we present the leading behavior of the first return probability density. This can be obtained naively by substituting equation~(\ref{ser2d}) in equation~(\ref{reca}). We obtain
\begin{equation}
\label{reca1}
\tilde F(0,s)\xrightarrow[s \rightarrow 0]{} 1+\frac{{\mathcal{D}}_{2d}}{D_{2d}}\frac{ \pi }{\log s}+...~,
\end{equation}
where $\mathcal{D}_{2d}$ is the effective diffusion constant in two dimensions defined in equation~(\ref{d2d}).
Performing a Laplace inversion of the above equation yields the leading behavior
\begin{eqnarray}
\label{reca3}
F(0,t)\xrightarrow[t \rightarrow \infty]{}\frac{{\mathcal{D}}_{2d}}{D_{2d}}\frac{ \pi}{t ~{\log^2 t}}=(1+2~ {\text{Pe}}^2)\frac{ \pi}{t~ {\log^2 t}}.
\end{eqnarray}
Once again, as in the one dimensional case, the long time properties of the first return are governed by the ratio of the modified diffusion constant $\mathcal{D}_{2d}$ and the intrinsic diffusion constant ${D}_{2d}$ which can be quantified in terms of the P\'eclet number $\text{Pe}=\epsilon/\sqrt{D_{2d} \gamma}$.
Similarly, the asymptotic behavior of the cumulative first return probability can be obtained as
\begin{equation}
\label{reca4}
R(0,t)\xrightarrow[t \rightarrow \infty]{} 1-\frac{{\mathcal{D}}_{2d}}{D_{2d}}\frac{\pi}{ {\log t}}+...~,
\end{equation}
 and the walk is recurrent with $R(0,\infty)=1$.
\subsection{Limiting cases}

Using the exact expression for the characteristic function of the probability of occupation of an arbitrary lattice site by a symmetric random walker, we derive the exact expression for the characteristic function of the probability of the first passage to any arbitrary lattice site.
Substituting~equations~(\ref{eq:x}) and (\ref{eq:y}) in equation~(\ref{reca0}) yields the exact analytic expression for the  characteristic function of the first passage time distribution of a symmetric random walker on a two dimensional infinite square lattice for any lattice site other than the origin as
\begin{small}
\begin{eqnarray}
\label{hf1}
\hspace{-2.5 cm}
{\tilde F(x,y,s)}_{srw}= \,
   _4F_3\biggl[\frac{1+|x|+|y|}{2},\frac{1+|x|+|y|}{2},\frac{2+|x|+|y|}{2},\frac{2+|x|+|y|}{2};1+|x|,1+|y|,\nonumber\\ \hspace{-2.5 cm} 1+|x|+|y|; {\left (\frac{4 D_{2d}}{s+4 D_{2d}} \right)}^2\biggr]/\left[ \left(\frac{s+4 D_{2d}}{ D_{2d}}\right)^{|x|+|y|} \frac{|x|! |y|!}{\pi (|x|+|y|)!}2 K\left[ {\left (\frac{4 D_{2d}}{s+4 D_{2d}} \right)}^2\right]\right].~
\end{eqnarray}
\end{small}
For diagonal sites $x=y$, we obtain the simplified expression
\begin{equation}
\label{hf}
{\tilde F(x,s)}_{srw}=\frac{\pi\Gamma \left(\frac{1}{2}+|x|\right) \, _2F_1\left[\frac{1}{2}+|x|,\frac{1}{2}+|x|;1+2
   |x|;{\left (\frac{4 D_{2d}}{s+4 D_{2d}} \right)}^2\right]}{ \left(\frac{s+4 D_{2d}}{2 D_{2d}}\right)^{2 |x|} \sqrt{\pi }~ \Gamma (1+|x|)2 K\left[{\left (\frac{4 D_{2d}}{s+4 D_{2d}} \right)}^2\right]},
\end{equation}
where ${\tilde F(x,s)}_{srw}={\tilde F(x,x,s)}_{srw}$. 
Similarly, we obtain the characteristic function of the probability of the first return to the origin ($x=y=0$) by substituting equation~(\ref{eq:y}) in equation~(\ref{reca}) as
\begin{equation}
\label{eq:z}
{\tilde F(0,s)}_{srw}=1-\frac{\pi }{2 K\left[{\left (\frac{4 D_{2d}}{s+4 D_{2d}} \right)}^2\right]}.
\end{equation}
\begin{figure} [t!]
\hspace{-1 cm}
 \includegraphics[width=1.15\linewidth]{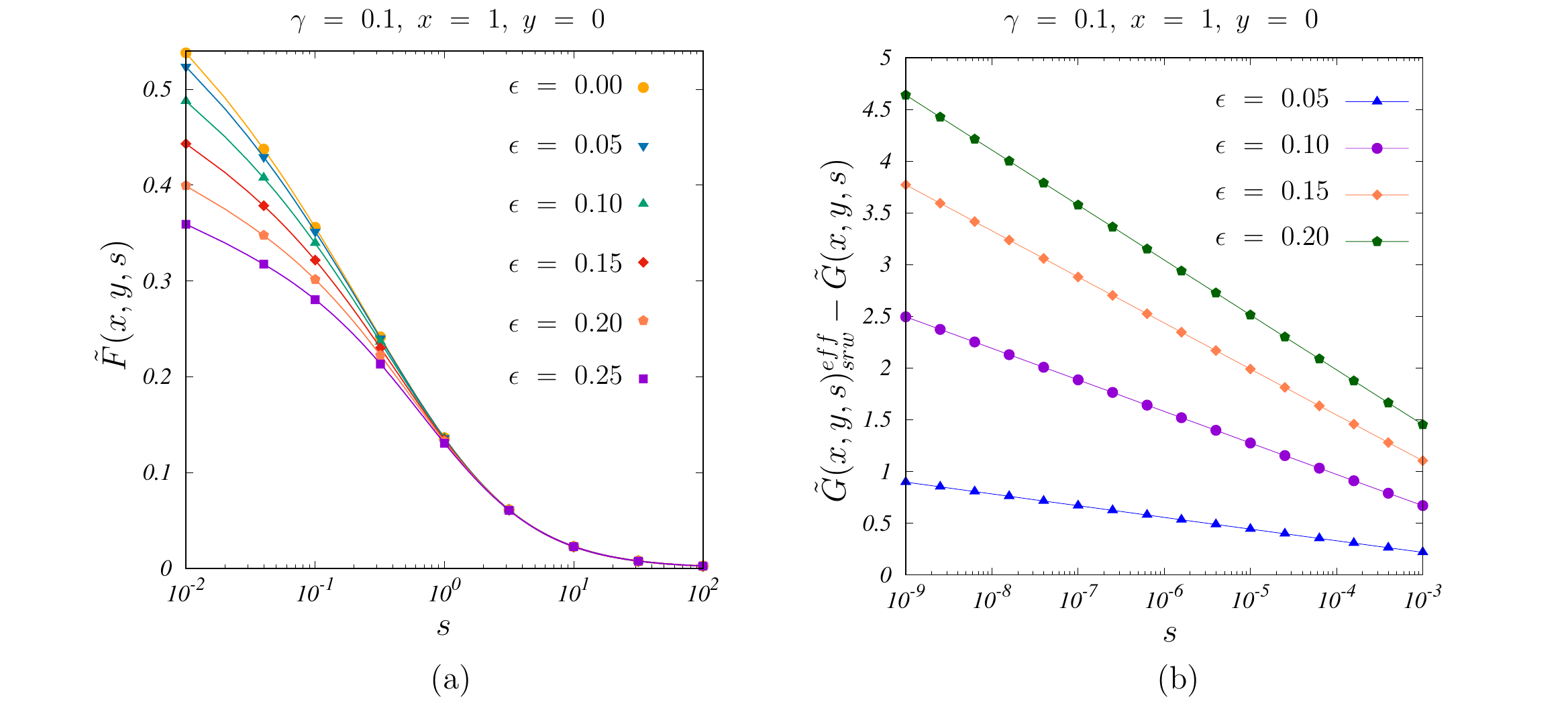} 
\caption{(a)~The Laplace transform of the first passage probability distribution of an active
random walker  $\tilde F(x,y,s)$, obtained from kinetic Monte Carlo simulations (points) plotted as
a function of $s$ for the lattice site $x = 1,~ y = 0$, for different values of $\epsilon$. The simulation data is averaged over $10^6$ realizations. We have done a numerical Laplace transform of simulation data in the time domain to obtain the characteristic function. The solid curves for non-zero $\epsilon$ correspond to the theoretical result in equation~(\ref{reca0}) along with the expression for $\tilde P(x,y,s)$ in equation~(\ref{eq:fourier_inv}). We
have performed a numerical Fourier inversion of the exact expression for  $\tilde P(k_x,k_y,s)$) provided in equation~(\ref{pks2dn}) to obtain $\tilde P(x,y,s)$). The solid curve for $\epsilon = 0.00$ which in turn corresponds to the limit of a symmetric random walk is obtained from the exact analytic expression in equation~(\ref{hf1}).~(b)~The difference between the Laplace transforms of the first passage probability distribution of a symmetric random walker with a modified diffusion constant~${\tilde G(x,y,s)}^{eff}_{srw} \equiv {[1-{\tilde F(x,y,s)}^{eff}_{srw}]}^{-1} $, and an active random walker $\tilde G(x,y,s) \equiv {[1-{\tilde F(x,y,s)}]}^{-1}$ plotted (lines with points) as a function of $s$ for the lattice site $x=1,~y=0$, for different values of $\epsilon$. For small $s$, the difference between these quantities varies at $O(-\log {s})$. The fixed parameter values used are $\gamma=0.1$ and $D_{2d}=0.25$.}
\label{fig:fxys_fxys_conv}
\end{figure}

In figure~\ref{fig:fxys_fxys_conv}(a), we display a plot of the Laplace transform of the first passage probability distribution of an active random walk $\tilde F(x,y,s)$, obtained from kinetic Monte Carlo simulations for different bias values $\epsilon$. We have performed a numerical Laplace transform of the simulation data for $F(x,y,t)$ to obtain $\tilde F(x,y,s)$. The solid curves plotted in the figure for non-zero $\epsilon$ correspond to the theoretical result in equation~(\ref{reca0}) along with the expression for $\tilde P(x,y,s)$ in equation~(\ref{eq:fourier_inv}). We have performed a numerical Fourier inversion of the exact expression for $\tilde P(k_x,k_y,s)$ provided in equation~(\ref{pks2dn}) to obtain $\tilde P(x,y,s)$. We show that in the $\epsilon \rightarrow 0$ limit, the simulation results converge to the exact expression for the characteristic function of the first passage probability distribution of a symmetric random walker ${\tilde F(x,y,s)}_{srw}$ provided in equation~(\ref{hf1}). In order to study the deviation of an active random walk from a symmetric random walk with a modified diffusion constant at large times, we next analyze the difference between the first passage distributions in Laplace space for small $s$ values.

\begin{figure}[t!]

        \centering
        \includegraphics[width=0.8\textwidth]{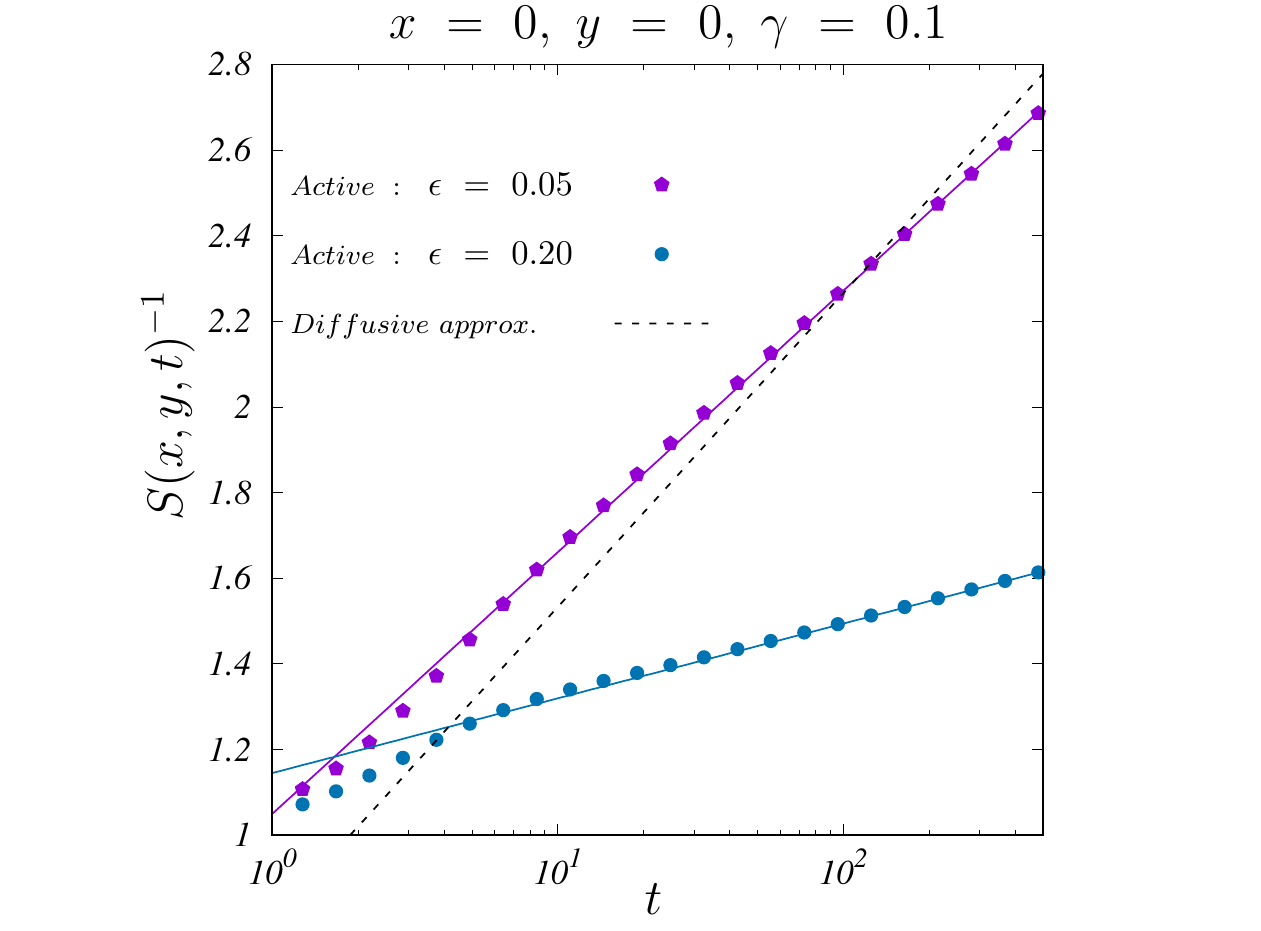}
\caption{(b) The survival probability $S(x,y,t)$, of an active random walker in two dimensions obtained from kinetic Monte Carlo simulations (points) plotted as a function of time for the lattice site $x=y=0$, for two different values of $\epsilon$. The solid curves correspond to the theoretical result in equation~(\ref{reca5}) shifted by an arbitrary constant. The dashed curve corresponds to an ordinary two dimensional Brownian motion with an effective diffusion constant $\mathcal{D}_{2d}$. For an ordinary Brownian motion in two dimensions, the slope of the curve is the same for any finite value of the diffusion constant in the asymptotic limit. The fixed parameter values used are $\gamma=0.1$ and $D_{2d}=0.25$. The simulation data is averaged over $10^6$ realizations.}\label{fig_r0t_2d}
\end{figure}


Based on our insight obtained from equation~(\ref{reca1}) for the first return probability to the origin $x=y=0$, we expect the quantity $[1-\tilde F(x,y,s)  ]$, to vary as $-1/\log s$ for an arbitrary lattice site ($x,y$), in the small $s$ regime. Hence, it would be easy to analyze the quantity ${[1-\tilde F(x,y,s)]}^{-1}$ denoted as $\tilde G(x,y,s)$ that decays as $-\log s$. We display a plot for the difference between these quantities for a symmetric random walk $\tilde G(x,y,s)^{eff}_{srw}$, with a modified diffusion constant and an active random walk $\tilde G(x,y,s)$, in the small $s$ limit for the lattice site $x=1,~y=0$, in figure~\ref{fig:fxys_fxys_conv}(b). Here, we obtain ${\tilde F(x,y,s)}^{eff}_{srw} $ by simply substituting $D_{2d}= \mathcal{D}_{2d}$ in equation~(\ref{hf1}) which yields ${\tilde G(x,y,s)}^{eff}_{srw} \equiv {[1-{\tilde F(x,y,s)}^{eff}_{srw}]}^{-1} $. For small $s$, the difference between these quantities also varies at $O(-\log{s})$. This implies that the first passage time densities of an active random walk and a symmetric random walk with a modified diffusion constant have distinct characteristic functions at the leading order. 

To test the above theoretical predictions, we also perform kinetic Monte Carlo simulations of an active random walk in two dimensions and study the deviation from a symmetric random walk at large times in terms of first passage probabilities. In figure~\ref{fig_r0t_2d}, we compare our theoretical prediction for the first return probability density of an active random walk in two dimensions with simulation results for the lattice site $x=y=0$. For convenience, we have plotted the inverse of the survival probability $S(x,y,t)$. We define the survival probability $S(x,y,t)$, as the probability that a RTP starting from the origin $x=y=0$ does not cross the lattice site $(x,y)$, up to time $t$. The survival probability is related to the first passage probability through
\begin{equation}
    f(x,y,t)=-\frac{\partial S(x,y,t)}{\partial t}.
\end{equation}
For $x=y=0$, the survival probability $S(0,0,t) \equiv S(0,t)$ is the probability that a RTP starting from the origin does not return to the origin up to time $t$. The survival probability and the cumulative first return probability provided in provided in equation~(\ref{reca4}) are related through, $S(0,t)=1-R(0,t)$. From~equation~(\ref{reca4}), for an active random walk in two dimensions, we thus obtain
\begin{equation}
\label{reca5}
\lim_{t \rightarrow \infty}S(0,t)^{-1}= \frac{{D}_{2d}}{{\mathcal{D}}_{2d}}\frac{\log t}{ \pi}+...~.
\end{equation}
The coefficient ${D}_{2d}/({{\mathcal{D}}_{2d}\pi})$ appearing in the RHS of the above equation appears as the slopes of the solid curves in the semi-log plot displayed in figure~\ref{fig_r0t_2d}. The result in equation~(\ref{reca5}) differs from the result within an effective diffusive approximation which yields a slope $1/\pi$, independent of the diffusion constants in the large time limit.


\section{Conclusions}
\label{sec:conclusions}
In this paper, we have investigated the first passage properties of active continuous time random walks with nearest neighbor jumps on one  and two dimensional infinite lattices. We focused on the simplest case where the waiting times are Poisson distributed. First, we derived exact expressions for the characteristic functions of the occupation probabilities of an active random walk. We analyzed the small and large time properties of the occupation probability and showed that at large times, the occupation probability of an active random walker resembles that of a symmetric random walker with a modified diffusion constant, validating previous findings in the literature. Additionally, we demonstrated that the subleading corrections to the occupation probability at intermediate times are not governed by the same modified diffusion constant. 

Using the exact expressions for the characteristic functions of the occupation probabilities, we studied the small and large time properties of the first passage time distributions of an active random walker. We showed that at large times, the first passage probabilities decay as $t^{-3/2}$ in one dimension and ${ 1}/{(t~ {\log^2 t})}$ in two dimensions just as in an ordinary Brownian motion. However, the asymptotic behavior of the first passage times is not governed by just an effective diffusion constant, and it depends crucially on the P\'eclet number unlike the case of occupation probabilities. We demonstrated  that at large times, activity increases the probabilities of the first passage to lattice sites close enough to the origin  and reduces the probabilities of the first passage to lattice sites far enough from the origin. Additionally, we derived the
first passage distributions of a biased random walker and a symmetric random walker as limiting cases. It would also be interesting to extend this model to higher dimensions, where less exact results are known.

\section{Acknowledgments}
I would like to thank Kabir Ramola for discussions and insightful suggestions.
I am grateful to Mustansir Barma for his careful reading of the manuscript and useful comments. I thank Prasad Perlekar, Dipanjan Mandal, Roshan Maharana, Vishnu V.~Krishnan, Pappu Acharya, Debankur Das, and Soham Mukhopadhyay for useful discussions. This project was funded by intramural funds at TIFR Hyderabad from the Department of Atomic Energy (DAE).


\section*{Appendix}
\appendix
\section{Laplace transform of the occupation probability of the origin in one dimension}\label{appendix_a0}
From~equation~(\ref{fl_inv}), we obtain the Laplace transform of the occupation probability of the origin in one dimension as
\begin{equation}
\label{fl_inv_origin}
\tilde P(0,s)=\frac{1}{2 \pi}\int_{-\pi}^{\pi}dk\frac{1}{(s+4D_{1d}\sin^2 \frac{k}{2})+\frac{4 \epsilon^2 \sin^2 k}{(s+4D_{1d}\sin^2 \frac{k}{2}+2 \gamma)}}.
\end{equation}
For simplicity, we substitute $D_{1d}=1/2$ in the following calculations and later replace the final expression with arbitrary diffusion constant $D_{1d}$.
Using the substitution $z=e^{ik}$ and equating $D_{1d}=1/2$ in the above equation yields
\begin{equation} 
\tilde P(0,s)=
\frac{i}{\pi(1-4 \epsilon^2)}\oint \frac{(z^2-2(1+s+2 \gamma)z+1)}{(z-z_1)(z-z_2)(z-z_3)(z-z_3)}dz,
\end{equation}
which is an integral over a unit circle in complex plane and $z_1~,z_2,~z_3$ and $z_4$ are the poles of the integrand $\tilde P(0,s)$. The explicit forms of the poles are given as
\begin{equation}
\label{eq:a16c}
z_{1,2}=\frac{-(1+s+\gamma)+\tilde f(s)\pm \sqrt{\tilde g(s)-2 \tilde f(s) (1+s+\gamma)}}{4 \epsilon ^2-1},
\end{equation}
\begin{equation}
\label{eq:a19c}
z_{3,4}=\frac{-(1+s+\gamma)-\tilde f(s) \pm \sqrt{\tilde g(s)-2 \tilde f(s) (1+s+\gamma)}}{4 \epsilon ^2-1},
\end{equation}
with
\begin{eqnarray} 
\tilde f(s)&=&
\sqrt{\gamma ^2+4 \epsilon ^2 [2 \gamma  (s+1)+s (s+2)]+16 \epsilon ^4},\nonumber\\
\tilde g(s)&=&
8 (\gamma +1) \left(\gamma +4 \epsilon ^2\right)+4 s^2 \left(4 \epsilon ^2+1\right)
+8 (\gamma +1) s \left(4 \epsilon ^2+1\right),\nonumber\\
\tilde h(s)&=&\sqrt{s (s+2) (s+2 \gamma ) (s+2+2 \gamma)}.
\label{fs_hs}
\end{eqnarray} 
The poles, $z_1$ and $z_3$ lie within the contour and thus we obtain the Laplace transform of the occupation probability of the origin as
\begin{equation}
\tilde P (0,s)= \frac{i}{\pi(1-4 \epsilon^2)} 2 \pi i \sum_{i=1,3} \text{Residues},
\end{equation}
which on simplification yields 
\begin{equation}
\label{rtp_d1_2}
\tilde P(0,s)=
\sqrt{\frac{\gamma ^2+4 \epsilon ^2 \gamma  (s+1) +2 \epsilon ^2 \left[\tilde h(s)+s (s+2)\right]}{s (s+2) \tilde f(s)^2}}.
\end{equation}
The expressions for the functions $\tilde f(s),~\tilde h(s)$ appearing in the above equation is provided in~equation~(\ref{fs_hs}).
In the same way, one can compute the Laplace transform of the occupation probability $\tilde P(x,s)$, for any arbitrary lattice site $x$, using the contour integration 
\begin{equation}
\tilde P(x,s)=
\frac{i}{\pi(1-4 \epsilon^2)}\oint \frac{z^{|x|}(z^2-2(1+s+2\gamma)z+1)}{(z-z_1)(z-z_2)(z-z_3)(z-z_3)}dz.
\end{equation}

To obtain the Laplace transform of the occupation probability of the origin for a process with arbitrary diffusion constant $D_{1d}$, we divide the RHS of~equation~(\ref{rtp_d1_2}) with $2 D_{1d}$ and rescale the bias and flipping rates as $\epsilon \rightarrow \frac{\epsilon}{2 D_{1d}}$ and $\gamma \rightarrow \frac{\gamma}{2 D_{1d}}$. This yields the exact expression in~equation~(\ref{rtp}).
\section{Survival probability of an active particle in one dimension - continuous space}\label{appendix_b}
\begin{figure} [h]
\centering
 \includegraphics[width=0.9\linewidth]{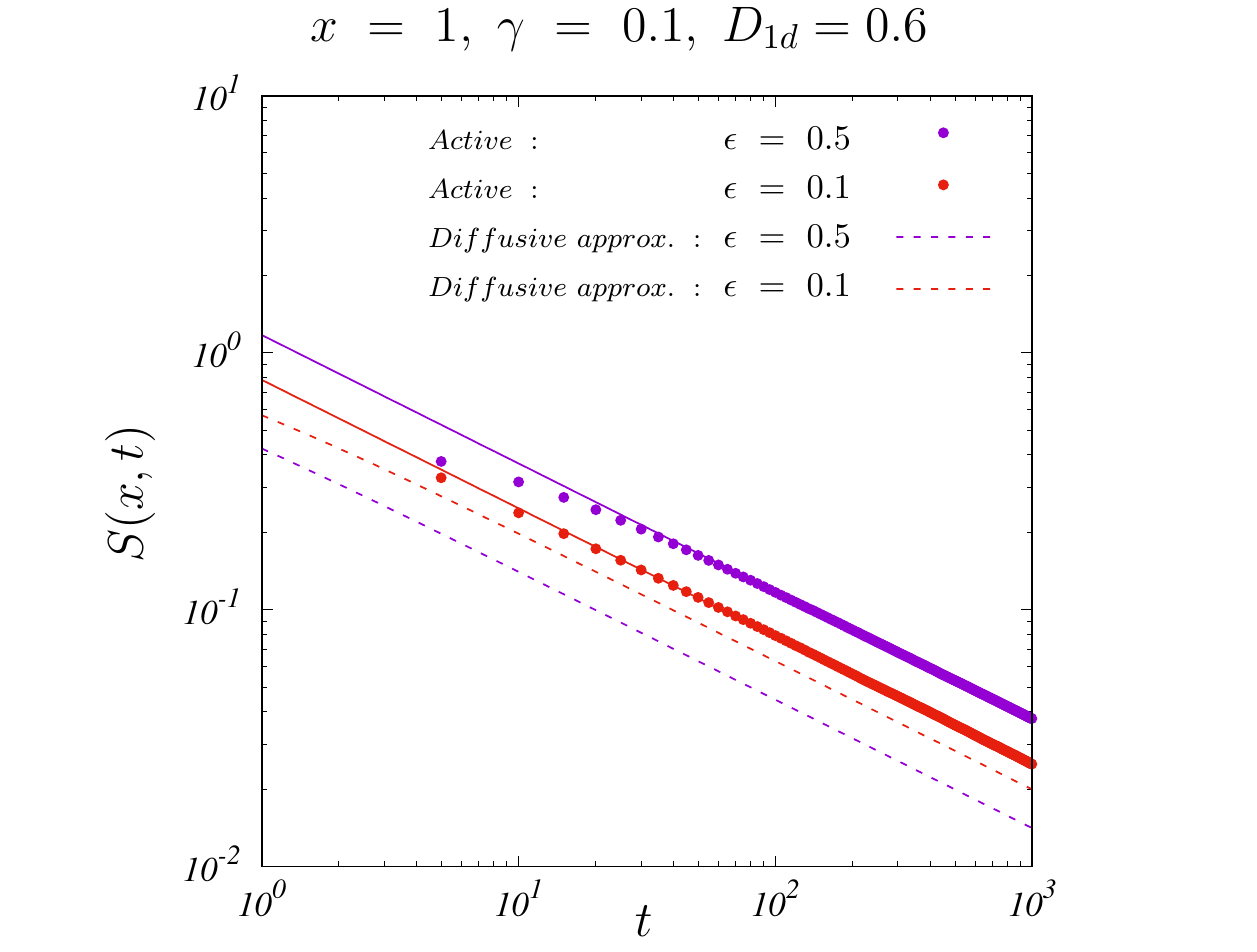}
 \caption{Continuous space simulation results (points) for the survival probability $S(x,t)$, of a RTP in one dimension for the lattice site $x=1$, plotted as a function of time for different values of $\epsilon$. The solid curves correspond to the analytic expression provided in~equation~(\ref{sxt_cont_exp}). The dashed curves correspond to an ordinary Brownian motion with an effective diffusion constant~$\mathcal{D}_{1d}$. The fixed parameter values used are $\gamma=0.1$, and $D_{1d}=0.6$. }\label{sxt_cont}
\end{figure} 
We study the motion of an active particle starting from the origin $x=0$, at time $t=0$ in one dimension following the Langevin equation,
\begin{equation}
\frac{dx}{dt}=2 \epsilon~\Sigma (t) + \sqrt {2 D_{1d}}~\Gamma (t).
\label{langevin}
\end{equation}
Here, $\Sigma (t)$ is a random variable that can switch values between $\pm 1$ at a Poisson rate $\gamma$ and $\Gamma (t)$ is a Gaussian white noise with mean zero and delta correlation in time~\cite{malakar2018steady},
\begin{equation}
  \braket {\Gamma(t)} =0,~  \braket {\Gamma(t) \Gamma (t')} =\delta (t-t').
\end{equation}
For an active particle in continuous space following the dynamics given in~equation~(\ref{langevin}), the first passage time density has the following asymptotic behavior
\begin{equation} 
\label{fxt_cont}
F(x,t)\xrightarrow[t \rightarrow \infty]{}
\left (\frac{\left| x\right|}{\sqrt{4 \pi \mathcal{D}_{1d}}}+ \bar{\xi}_\frac{3}{2} \right)\frac{1}{t^{\frac{3}{2}}},
\end{equation}
where the coefficient $\bar{\xi}_\frac{3}{2}$ has the explicit form
\begin{equation}
\label{xi_co}
\bar{\xi}_\frac{3}{2}=    \frac{\sqrt{\frac{2}{\gamma }} \epsilon ^2}{\sqrt{4 \pi } \gamma  \mathcal D_{1d}} \left(1-e^{-\frac{x
   \sqrt{2 D_{1d} \gamma +4 \epsilon ^2}}{D_{1d}}}\right).
\end{equation}
For the lattice sites far enough from the origin ($x \rightarrow \infty$), the coefficient~$\bar{\xi}_\frac{3}{2}$ reduces to the constant,
\begin{equation}
\label{xi_cont}
\lim_{x \rightarrow \infty}\bar{\xi}_\frac{3}{2}=\bar{\xi}^c_\frac{3}{2}= \frac{\sqrt{\frac{2}{\gamma }} \epsilon ^2}{\sqrt{4 \pi } \gamma  \mathcal{D}_{1d}}.
\end{equation}
Equation~(\ref{fxt_cont}) can be derived by solving the associated Fokker-Planck equations for the survival probability in continuous space~\cite{malakar2018steady} or by taking a continuum limit~\cite{PhysRevE.105.064103} of the corresponding expression for the lattice model provided in~equation~(\ref{series_fxt_actual}). The long time behavior of the survival probability $S(x,t)$ can be obtained by integrating the expression provided in~equation~(\ref{fxt_cont}) in time and this yields
\begin{equation} 
\label{sxt_cont_exp}
S(x,t)\xrightarrow[t \rightarrow \infty]{}
\left (\frac{\left| x\right|}{\sqrt{ \pi \mathcal{D}_{1d}}}+ 2 \bar{\xi}_\frac{3}{2} \right)\frac{1}{t^{\frac{1}{2}}}.
\end{equation}
In figure~\ref{sxt_cont}, we display a plot comparing the continuous space Monte Carlo simulation results for the long time behavior of the survival probability and the analytic expression provided in~equation~(\ref{sxt_cont_exp}) for the lattice site $x=1$. We infer that the effective Brownian approximation does not capture the right behavior of the first passage probabilities at large times in continuous space.

The expression provided in~equation~(\ref{sxt_cont_exp}) resembles the expression for the survival probability of a RTP derived in~\cite{le2019noncrossing} for the continuous space model. However, in that case, the diffusion rate $D_{1d}=0$. In the limit of zero diffusion, $\bar{\xi}_\frac{3}{2}$ provided in equation~(\ref{xi_co}) reduces to the constant $\bar{\xi}^c_\frac{3}{2}(D_{1d}=0)$ given as
\begin{equation}
\bar{\xi}^c_\frac{3}{2}(D_{1d}=0)= \frac{\sqrt{\frac{2}{\gamma }} \epsilon ^2}{\sqrt{4 \pi } \gamma  \mathcal{D}^0_{1d}}=\frac{1}{\sqrt{4 \pi \mathcal{D}^0_{1d} }}\bar{\xi}_{\text{Milne}},
\end{equation}
where $\mathcal{D}^0_{1d}=2\epsilon^2/\gamma$ is the effective diffusion constant for the zero diffusive case and $\bar{\xi}_{\text{Milne}}=\frac{\epsilon}{\gamma}$ is the Milne extrapolation length.
Thus, we obtain

\begin{equation} 
\label{sxt_cont_exp_zero_dif}
S(x,t)\xrightarrow[t \rightarrow \infty,~ D_{1d}=0]{}
\frac{1}{\sqrt{ \pi \mathcal{D}^0_{1d}t}} \left ( \left | x \right |+\bar{\xi}_{\text{Milne}} \right ).
\end{equation}
This expression is exactly the expression for the asymptotic behavior of the survival probability derived in~\cite{le2019noncrossing}. We notice that for the zero diffusive case, the correction to the survival probability due to activity $\bar{\xi}_{\text{Milne}}$ is independent of the space variable $x$ and is a constant dependent only on the rates. However, any finite value of diffusion introduces non trivial correction to the survival probability as provided in equation~(\ref{xi_co}).
\newpage
\section{List of coefficients - one dimension}\label{appendix_a}
\begin{table}[b]
\caption{\label{table_arw}List of coefficients appearing in the expressions for various quantities related to an active random walk in one dimension.}
\begin{tabular}{@{} l l l }
\br
Quantity & Coefficient & Expression \\ \mr
\multirow{4}{*}{$P(0,t)~(t \rightarrow 0)$} & $\tilde{\beta}_{0}$ & $1$ \\
 & $\tilde{\beta}_{1}$  & $-2D_{1d}$ \\
 & $\tilde{\beta}_{2}$ & $ 2 {D_{1 d}}^2+\eta$ \\
 & $\tilde{\beta}_{3}$  & $-\frac{2}{3} \left(2 {D_{1 d}}^3+3 D_{1 d} \eta -\gamma  \epsilon ^2\right)$ \\ 
\mr
\multirow{3}{*}{$F(0,t)~(t \rightarrow 0)$} & $\tilde{\rho}_1$ & $2 \eta $ \\
 & $\tilde{\rho}_2$  & $-4D_{1d}\eta+2 \gamma \epsilon^2$ \\
 & $\tilde{\rho}_3$ & $ \frac{1}{3}   \left[\eta (12 {D_{1 d}}^2+\eta)-4 \gamma \epsilon^2 (\gamma+3 D_{1d}) \right]$ \\
\mr
\multirow{4}{*}{$ P(0,t)~(t \rightarrow \infty)$} & $\tilde{\alpha}_{\frac{1}{2}}$ & $\frac{1}{\sqrt{4 \pi \mathcal D_{1d}}}$ \\
& $\tilde{\alpha}_{1}$ & $2 \epsilon ^2 \sqrt{\frac{1+\gamma }{\left(\gamma +4 \epsilon ^2\right)^3}}$ \\
 & $\tilde{\alpha}_{\frac{3}{2}}$ & $\frac{D_{1 d} \gamma ^2+8  \left(D_{1 d}+\gamma \right) \epsilon ^2-8 \epsilon ^4/\gamma}{ 16 {\mathcal D_{1d}}^2\sqrt{4\pi \mathcal D_{1d}  }}$ \\
 & $\tilde{\alpha}_{2}$ & $\frac{\epsilon ^2 \left(-\gamma  (1+2 \gamma )+8 (2+\gamma  (4+3 \gamma )) \epsilon ^2+16 \gamma  \epsilon ^4\right)}{2 \sqrt{1+\gamma } \left(\gamma +4 \epsilon ^2\right)^{7/2}}$ \\
  \mr
\multirow{2}{*}{$P(x,t)~(t \rightarrow \infty)$} & $\tilde{\sigma}_{\frac{1}{2}}$ & $\frac{1}{\sqrt{4 \pi \mathcal D_{1d}}}$ \\
 & $\tilde{\sigma_{1}}$ & \pbox{20 cm}{$\frac{1}{2}\frac{\sqrt{\frac{1}{D_{1d}}+\frac{2}{\gamma }} \epsilon ^2}{ \gamma  {\mathcal{D}_{1d}}^{\frac{3}{2}}}~ \times$\\$  \left(\frac{{D_{1d}}({D_{1d}}+ \gamma) +\epsilon ^2-\sqrt{{D_{1d}} (2 {D_{1d}}+\gamma ) \gamma \mathcal D_{1d}}}{\eta}\right)^{|x|}$} \\
\mr
\multirow{5}{*}{$ F(0,t)~(t \rightarrow \infty)$} & $\tilde{\phi}_1$ & $1$ \\
 & $\tilde{\phi}_\frac{3}{2}$ & $\frac{\mathcal D_{1d}}{D_{1d}}\frac{1}{\sqrt{4 \pi \mathcal D_{1d}}} $ \\
 & $\tilde{\phi}_2$ & $\frac{\epsilon ^2 }{{(\gamma D_{1d})}^\frac{3}{2}}\sqrt{\frac{\gamma +2 D_{1d}}{\mathcal D_{1d}}}$ \\
  & $\tilde{\phi}_\frac{5}{2}$ & $3\frac{\left[ 3+8 \epsilon ^2 \left(\gamma -D_{1d}\right)( \gamma D_{1d}+ \epsilon ^2 )/\left({D_{1d}}^2\gamma^3\right) \right]}{16 \mathcal D_{1d} \sqrt{4\pi \mathcal D_{1d}} }$ \\
\mr
\multirow{1}{*}{$F(x,t)~(t \rightarrow \infty)$} & $\tilde{\xi}_{1}$ & $1$ \\
& $\tilde{\xi}_{\frac{3}{2}}$ & \pbox{20 cm}{$\frac{\sqrt{\frac{1}{D_{1d}}+\frac{2}{\gamma }} \epsilon ^2}{\sqrt{4 \pi } \gamma  \mathcal{D}_{1d}}~ \times$\\$  \left[1-\left(\frac{{D_{1d}}({D_{1d}}+ \gamma) +\epsilon ^2-\sqrt{{D_{1d}} (2 {D_{1d}}+\gamma ) \gamma \mathcal D_{1d}}}{\eta}\right)^{|x|}\right]$} \\
\br
\end{tabular}
\end{table}
\begin{table}
\caption{\label{table_brw}List of coefficients appearing in the expressions for various quantities related to a passive random walk in one dimension.}
\begin{tabular}{@{} l l l }
\br
Quantity & Coefficient & Expression \\ 
\mr
\multirow{4}{*}{${P(0,t)}_{brw}~(t \rightarrow 0)$} & $\beta_0$ & $1$ \\
 & $\beta_1$  & $-2D_{1d}$ \\
 & $\beta_2$ & $ 2 {D_{1 d}}^2+\eta$ \\
 & $\beta_3$  & $-\frac{2}{3}\left(2 {D_{1 d}}^3+3 D_{1 d} \eta \right)$ \\ 
\mr
\multirow{3}{*}{${F(0,t)}_{brw}~(t \rightarrow 0)$} & $\rho_1$ & $2 \eta$ \\
 & $\rho_2$  & $-4D_{1d}\eta $ \\
 & $\rho_3$ & $\frac{1}{3} \eta  \left(12 {D_{1 d}}^2+\eta \right)$ \\
\mr
\multirow{3}{*}{$ {P(0,t)}_{srw}~(t \rightarrow \infty)$} & $\alpha_\frac{1}{2}$ & $\frac{1}{\sqrt{4\pi D_{1d}  }}$ \\
 & $\alpha_\frac{3}{2}$  & $\frac{1}{ 16 D_{1d}\sqrt{4\pi D_{1d}  }}$ \\
 & $\alpha_\frac{5}{2}$ & $\frac{9}{512 {D_{1d}}^2 \sqrt{4\pi D_{1d}  }}$ \\
\mr
\multirow{3}{*}{$ {P(x,t)}_{srw}~(t \rightarrow \infty)$} & $\sigma_\frac{1}{2}$ & $\frac{1}{\sqrt{4\pi D_{1d}  }}$ \\
 & $\sigma_\frac{3}{2}$  & $\frac{1}{ 16 D_{1d}\sqrt{4\pi D_{1d}  }}$ \\
 & $\sigma_2$ & $\frac{{|x|}}{48 {D_{1d}}^2 }$ \\
\mr
\multirow{3}{*}{$ {F(0,t)}_{srw}~(t \rightarrow \infty)$} & $\phi_\frac{3}{2}$ & $\frac{1}{\sqrt{4\pi D_{1d}  }}$ \\
 & $\phi_\frac{5}{2}$  & $\frac{9}{16 D_{1 d}  \sqrt{4 \pi D_{1 d}} }$ \\
 & $\phi_\frac{7}{2}$ & $\frac{345 }{512 D_{1 d}^2 \sqrt{4 \pi D_{1 d}} }$ \\
 \mr
\multirow{3}{*}{$ {F(x,t)}_{srw}~(t \rightarrow \infty)$} & $\xi_1$ & $1$ \\
 & $\xi_\frac{5}{2}$  & $\frac{{|x|}}{ 16 D_{1d}\sqrt{4\pi D_{1d}  }}$ \\
\br
\end{tabular}
\end{table}
\section*{References}
\bibliographystyle{unsrt}
\bibliography{bibtex.bib}

\end{document}